\documentclass[lettersize,journal]{IEEEtran}

\usepackage{amsmath,amsfonts}
\usepackage{algorithmic}
\usepackage{algorithm,algorithmic}
\usepackage{array}
\usepackage[caption=false,font=normalsize,labelfont=sf,textfont=sf]{subfig}
\usepackage{textcomp}
\usepackage{stfloats}
\usepackage{url}
\usepackage{verbatim}
\usepackage{graphicx}
\usepackage{cite}
\usepackage{amsmath,graphicx}
\usepackage{bm}
\usepackage{cite,color}
\usepackage{amsmath,amssymb,amsfonts}
\usepackage{algorithmic}
\usepackage{algorithm,algorithmic}
\usepackage{amsthm}
\usepackage{multirow}
\usepackage{booktabs}
\usepackage{hyperref}

\theoremstyle{definition}
\newtheorem{theorem}{Theorem}
\newtheorem{remark}{Remark}
\newtheorem{proposition}{Proposition}

\def\hbf{\mathbf{h}}

\def\fbf{\mathbf{f}}

\begin{document}

\title{Power-Efficient Over-the-Air Aggregation with Receive Beamforming for Federated Learning  }

\author{Faeze Moradi Kalarde, Min Dong,~\IEEEmembership{Fellow,~IEEE,} Ben Liang,~\IEEEmembership{Fellow,~IEEE,} \\Yahia A. Eldemerdash Ahmed, and Ho Ting Cheng 
\thanks{Part of this work was presented in \cite{Mswim}.
Faeze Moradi and Ben Liang are with the Department of ECE, University of Toronto, Toronto, ON M5S 1A1, Canada
(e-mail: faeze.moradi@mail.utoronto.ca; liang@ece.utoronto.ca).
Min Dong is with the Department of ECSE, Ontario Tech University, Oshawa, ON L1G 0C5, Canada (e-mail:
min.dong@ontariotechu.ca). Yahia Ahmed and Ho Ting Cheng are with Ericsson Canada, Ottawa, K2K 2V6, Canada (e-mail:
yahia.ahmed@ericsson.com; ho.ting.cheng@ericsson.com).}}

\maketitle

\begin{abstract}
This paper studies power-efficient uplink transmission design for federated learning (FL) that employs over-the-air analog aggregation and multi-antenna beamforming at the server. We jointly optimize device transmit weights and receive beamforming at each FL communication round to minimize the total device transmit power while ensuring convergence in FL training. Through our convergence analysis, we establish sufficient conditions on the aggregation error to guarantee FL training convergence. Utilizing these conditions, we reformulate the power minimization problem into a unique bi-convex structure that contains a transmit beamforming optimization subproblem and a receive beamforming feasibility subproblem. Despite this unconventional structure, we propose a novel alternating optimization approach that guarantees monotonic decrease of the objective value, to allow convergence to a partial optimum.
We further consider imperfect channel state information (CSI),
which requires accounting for the channel estimation errors in the power minimization problem and FL convergence analysis.
We propose a CSI-error-aware joint beamforming algorithm, which can substantially outperform one that does not account for channel estimation errors.
Simulation with canonical classification datasets demonstrates that our proposed methods achieve significant power reduction compared to existing benchmarks across a wide range of parameter settings, while attaining the same target accuracy under the same convergence rate.
\end{abstract}


\maketitle

\section{Introduction}
\IEEEPARstart{F}{ederated} learning (FL) leverages the computational capabilities of local devices to collaboratively train a global machine learning model while preserving the privacy of their local datasets \cite{mcmahan2017communication}. A central server coordinates the devices to perform local model updates using their local datasets and aggregates these updates for a global model update.
However, the information exchange between the edge devices and the server over wireless links can impose significant stress on the limited communication resources, including wireless bandwidth and device power. Conventional orthogonal multiple access methods may incur a large delay when the number of participating devices becomes large. To reduce the communication overhead, over-the-air aggregation using analog transmission has been proposed in \cite{firstyang2020federated, yang2020federated}.
In this approach, the devices transmit their model updates simultaneously using analog modulation 
over a shared multiple access channel, achieving natural model aggregation through superposition. This over-the-air computation technique has received significant interest due to its efficient utilization of bandwidth and reduced communication latency in contrast to conventional multiple access techniques over orthogonal channels, and various designs and analyses based on this technique have been proposed 
\cite{amiri2020machine,amiri2020federated,sery2020analog,guo2020analog, newone2}.

However, over-the-air computation is susceptible to aggregation errors, which propagate over the FL training iterations, thereby degrading the training performance. The aggregation errors are mainly caused by two sources: receiver noise and wireless channel distortion. To improve transmission quality and reduce aggregation errors, beamforming can be employed. Specifically, a multi-antenna server can utilize receive beamforming, while the devices can form distributed transmit beamforming to improve the transmission quality.
In designing beamforming schemes for FL, it is crucial to aim not only to achieve the desired learning accuracy but also to ensure power efficiency for devices. This will reduce energy consumption and minimize network interference. 
However, most existing uplink beamforming designs for FL with over-the-air aggregation focus on addressing the received signal quality for FL training convergence \cite{firstyang2020federated, yang2020federated, Paper2023, AliBereyhi, liu2021reconfigurable}, and none of them have considered power efficiency in their design solutions.
\subsection{Related Works} 
Recent studies have considered the transmit power efficiency in FL with over-the-air aggregation assuming a single-antenna server \cite{zhu2019broadband, sery2021over, wang2022online, ImperfectDownlink, ICCPaper2020,  GunduzPaper, newoneadded}. A common approach in these studies is to design the transmit weights of devices based on channel inversion while incorporating different types of device transmit power constraints. For example, to study FL learning performance, \cite{zhu2019broadband} and \cite{sery2021over} employ channel inversion to determine the device transmit weights and design a device selection scheme based on a channel strength threshold to meet the average transmit power limit at each device. 
In \cite{wang2022online}, a long-term transmit power constraint is imposed at each device, and the local model update is optimized to minimize the cumulative FL training loss.
Considering both noisy downlink and uplink transmission, \cite{ImperfectDownlink} aims to minimize the upper bound on the global loss optimality gap by jointly designing the device transmit weights and the server transmit power while considering constraints on either the individual or total uplink transmit power of devices during each communication round. 
In \cite{ICCPaper2020} and \cite{GunduzPaper}, energy-aware dynamic device selection algorithms are proposed to optimize FL training performance while confining the energy consumption of devices. In \cite{newoneadded}, the overall energy consumption of devices is minimized through the joint optimization of the transmission probability and the number of local computing iterations, while guaranteeing convergence.
However, all of these designs are limited to the single-antenna server setting.

For a multi-antenna server, it is shown in \cite{zhu2018mimo} that beamforming techniques can be employed to reduce the impact of noise and channel distortion in over-the-air computation. This study optimizes the receive beamforming to minimize the mean squared error (MSE) of the received aggregated signal subject to the average transmit power constraint at each device. 
Joint receive beamforming and device selection strategies are proposed in \cite{yang2020federated} and \cite{Paper2023} to maximize the number of selected devices while ensuring a bounded MSE at the server receiver and a bounded average transmit power at each device.
For the same problem, a low-complexity device selection algorithm is further proposed in \cite{AliBereyhi} using the method of matching pursuit. Joint receive beamforming and device selection algorithms are proposed in \cite{MyPaper} to maximize the convergence rate under a limited average power constraint at the devices. 
The uplink design for FL with the aid of a reconfigurable intelligent surface (RIS) is studied in \cite{liu2021reconfigurable} and \cite{jiang2019over}. In \cite{jiang2019over}, the receive beamforming and RIS reflective coefficients are jointly optimized to reduce the MSE of the received signal under an average transmit power constraint. Additionally, \cite{liu2021reconfigurable} considers device selection and jointly optimizes receive beamforming, device selection, and RIS reflective coefficients to improve the FL convergence rate subject to the average transmit power constraint at each device.
Also, an over-the-air aggregation method for simultaneously training multiple models is proposed in \cite{multimodelChong}, where receive beamforming and transmit power control are jointly optimized under the average transmit power constraint for the proposed multi-model aggregation method.
Besides uplink beamforming design, joint downlink and uplink beamforming design to maximize the FL training convergence is proposed in \cite{chongwiopt} subject to the server and device average transmit power constraints.
All these works focus on optimizing certain FL training metrics while imposing a transmit power constraint. In comparison, we aim to optimize the power efficiency for uplink transmission in FL with over-the-air aggregation.

Besides analog over-the-air aggregation, energy efficiency has been studied in digital transmission for FL uplink aggregation such as \cite{digit3WalidSaad} and\cite{digit5}, which focus on either minimizing or bounding device energy consumption. These proposed approaches are limited to digital transmission and are not applicable to the scenario considered in this work, which uses analog aggregation.

\subsection{Contribution}
This paper investigates power-efficient design for FL with over-the-air analog aggregation. Assuming a multi-antenna server, we aim to design uplink beamforming that maximizes the transmit power efficiency at the devices while ensuring FL training convergence. In particular, we jointly optimize the receive beamforming and device transmit weights for each FL communication round, assuming time-varying channel conditions. Our contribution can be summarized as follows:
\begin{itemize}
\item Based on gradient-based uplink aggregation, we formulate the joint beamforming problem to minimize
the total device transmit power while guaranteeing
FL training convergence to the optimal model. To tackle the challenging FL convergence constraint, we provide a set of sufficient conditions on the aggregation error that ensure convergence, which are then expressed in terms of the device transmit weights and receive beamforming. Our convergence analysis differs substantially from the existing methods for uplink beamforming design in FL. Instead of relying on the upper bounds of the optimality gap, which generally do not approach zero, our approach establishes conditions that guarantee convergence to the optimal model.

\item We reformulate the power minimization problem by replacing the convergence constraint with the derived sufficient conditions on the aggregation error for convergence. This leads to a unique bi-convex structure that contains a transmit beamforming \textit{optimization} subproblem and a receive beamforming \textit{feasibility} subproblem. We propose a novel alternating optimization approach, which utilizes the constraint function to structure the receive beamforming optimization, ensuring monotonic decrease of the objective value. Thus, our method guarantees convergence to a partial optimum solution. Although the transmit and receive beamforming subproblems are challenging to solve in their original forms, we show that both can be transformed into convex
quadratic programming (QP) problems, enabling efficient solutions. Moreover, we propose an effective initialization method to accelerate the convergence of the iterative procedure.

\item We further consider the scenario where only imperfect channel state information (CSI) is available at the server. In particular, we extend our developed framework to include estimated CSI and its associated errors in the analysis. We propose a CSI-error-aware joint beamforming algorithm that utilizes alternating optimization. Specifically, we show that the transmit beamforming subproblem can be transformed into a convex quadratically constrained quadratic programming (QCQP) problem, which can be solved efficiently.

\item Our experiments on the classification of canonical datasets over wireless networks demonstrate the power efficiency of our proposed methods. They lead to significant reduction in power consumption compared with the existing benchmarks for a wide range of parameter settings while attaining the same target accuracy under the same convergence rate.

\end{itemize}

\subsection{Organization}
The rest of this paper is organized as follows. Section II presents the system model and problem formulation. Section III describes the joint transmit and receive beamforming design, assuming perfect CSI. Section IV introduces the CSI-error-aware beamforming design. Simulation results are provided in Section V, and conclusions are presented in Section VI.

\allowdisplaybreaks
\section{System Model and Problem Formulation}

\subsection{FL System }
We consider a wireless network comprising a central server and $M$ edge devices. Each device, denoted by index $m$, contains a local training dataset of size $K_m$ represented by $\mathcal{D}_m=\{(\mathbf{x}_{m,k}, y_{m,k}): 1\le k \le K_m \}$, where $\mathbf{x}_{m,k}$ is the $k$-th data feature vector, and $ y_{m,k}$ is its corresponding label. The aim of the edge devices is to cooperatively train a global model on the server, capable of predicting the true labels of data feature vectors for all devices while ensuring the privacy of their local datasets. We define the empirical local training loss function for device $m$ as follows:
        \begin{align}\label{LocalLossDef}
            F_m(\mathbf{w}; \mathcal{D}_m)  \triangleq \frac{1}{K_m} \sum \limits_{k=1}^{K_m} l(\mathbf{w}; \mathbf{x}_{m,k} , y_{m,k}) ,
        \end{align}
where $\mathbf{w} \in \mathbb{R}^D$ is the global model parameter vector and $l(\cdot)$ is the sample-wise training loss associated with each data sample. Then, the global training loss function is
    \begin{align}\label{GlobalLossDef}
    F(\mathbf{w})= \frac{1}{K}\sum \limits_{m=1}^M K_m F_m(\mathbf{w}; \mathcal{D}_m ),
    \end{align}
where $K= \sum_{m} K_m $ is the total number of training samples over all devices. In this study, we adopt the conventional Federated Stochastic Gradient Descent (FedSGD) technique \cite{mcmahan2017communication} for iterative model training in FL, where the server updates the model parameters using an aggregation of the gradients derived from the local loss functions at all devices. The learning goal is to determine the optimal global model $\mathbf{w}^\star$ that minimizes the global training loss function $F(\mathbf{w})$. We refer to each cycle of the algorithm for a global model update as a communication round. The operation in communication round $t$ involves the following steps:
\begin{enumerate}
    \item \textbf{Downlink phase:} The server broadcasts the model parameter vector $\mathbf{w}_t$ to all devices.
    \item \textbf{Local gradient computation:} Each device $m$ computes the gradient of its local loss function, given by $\mathbf{g}_{m,t} \triangleq \nabla F_m(\mathbf{w}_t; \mathcal{D}_m) \in \mathbb{R}^D$,
    which is the gradient of $F_m(\cdot)$ at $\mathbf{w}_t$.
    \item \textbf{Uplink phase:} The devices transmit their local gradients to the server via the uplink wireless channels.

    \item \textbf{Model updating:} The server computes a weighted aggregation of the local gradients to update the global model. In an ideal scenario where the local gradients can be received accurately at the server, $\mathbf{r}_t \triangleq \sum_{m=1}^M K_m \mathbf{g}_{m,t}$ is utilized to update $\mathbf{w}_t$. However, in practical settings, only an approximation $\mathbf{\hat r}_t$ is feasible at the server due to the effects of wireless channels and receiver noise. Therefore, the server updates the global model as
    \begin{align}\label{ModelUpdate}
                \mathbf{w}_{t+1} = \mathbf{w}_t - \gamma_t \frac{\mathcal{R}({\mathbf{\hat r}_t})}{K},
    \end{align}
where $\gamma_t$ is the learning rate in round $t$ and $\mathcal{R}(\cdot)$ returns the real part of a complex variable.   
\end{enumerate}

\begin{figure}[t]
\centerline{\includegraphics[ width=0.4\textwidth, height=0.3\textwidth]{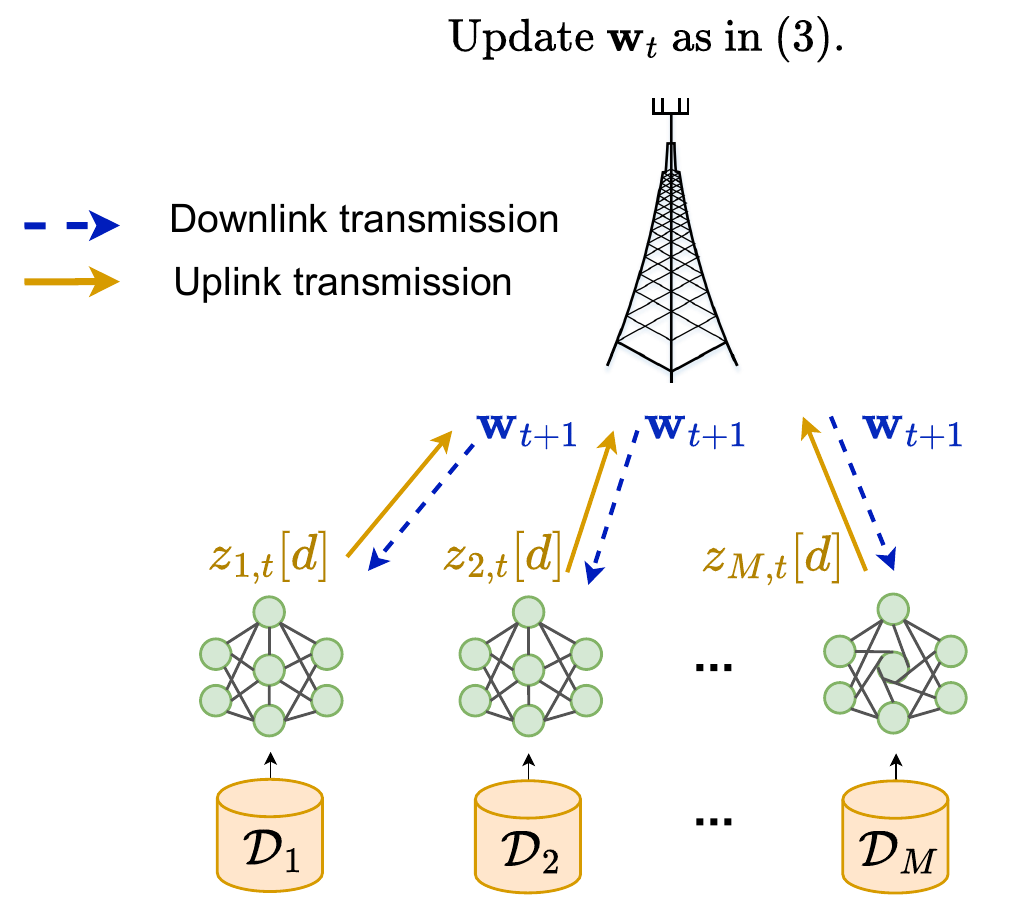}}
\caption{An illustration of federated learning with over-the-air aggregation.\label{systemmodel}}
\end{figure}

\subsection{FL with Over-the-Air Aggregation}
We assume that each device is equipped with a single antenna, while the server is equipped with $N$ antennas.
The wireless uplink channel between device $m$ and the server during communication round $t$ is represented by $\mathbf{h}_{m,t} \in \mathbb{C}^{N}$. In this work, we first propose our beamforming design assuming the server has perfect CSI\ knowledge of $\{\mathbf{h}_{m,t}: m=1,\ldots,K\}$ at the beginning of each round $t$. In Section \ref{impsection}, we extend our method to consider the beamforming design under the estimated CSI for  $\{\hat{\mathbf{h}}_{m,t}: m=1,\ldots,K\}$.

We utilize over-the-air computation for efficient aggregation of the local gradients at the server. This is achieved by analog aggregation over multiple access channels, as proposed in \cite{zhu2019broadband}. In each communication round $t$, the devices send their local gradients $\mathbf{g}_{m,t}$'s to the server simultaneously using the same frequency resource over $D$ time slots. Specifically, in time slot $d$, each device $m$ first normalizes the $d$-th entry $g_{m,t}[d]$ of $\mathbf{g}_{m,t}$ by $v_{m,t}= \frac{\|\mathbf{g}_{m,t}\|}{\sqrt{D}}$ and  then  sends it using the transmit weight $a_{m,t}\in\mathbb{C}$. The corresponding transmitted signal over $D$ time slots is denoted by $\mathbf{z}_{m,t}$, its $d$-th entry is given by
\begin{align}\label{TransmitEq}
    z_{m,t}[d] = a_{m,t}\frac{g_{m,t}[d]}{v_{m,t}}, d=1,\ldots,D.
\end{align}
As a result, the average transmit power per time slot is $\frac{\|\mathbf{z}_{m,t}\|^2}{D}=\left|a_{m,t}\right|^2$. We assume the average transmit power of each device is limited by $P_0$, i.e., $\left|a_{m,t}\right|^2 \le P_0$, $\forall m, \forall t$.
The corresponding received signal at the server in time slot $d$ of round $t$ is denoted by $\mathbf{y}_{d,t}$ and is given by
        \begin{align}\label{eq10}
            \mathbf{y}_{d,t} = \sum_{m=1}^M \mathbf{h}_{m,t} z_{m,t}[d]+\mathbf{n}_{d,t},
        \end{align}
where $\mathbf{n}_{d,t} \in \mathbb{C}^{N}$ is the receiver noise vector following the complex Gaussian distribution as $\mathcal{CN}(\mathbf{0}, \sigma_n^2\mathbf{I})$ with $\sigma_n^2$ being the noise variance, and $\mathbf{n}_{d,t}$ is independently and
identically distributed (i.i.d.) over $t$ and $d$. To facilitate the server to recover the local gradient from its normalized version, each device $m$ also sends the normalization scalar $v_{m,t}$ to the server in each communication round using a separate digital signaling channel, assuming error-free reception. 

The server applies receive beamforming to process the received signal. Let $\mathbf{f}_t \in \mathbb{C}^N$ denote the receive beamforming vector at round $t$. The post-processed received signal in time slot $d$ of round $t$ is given by
    \begin{equation}\label{RHatDef}
  \hat{r}_t[d]=  \mathbf{f}_t^H \mathbf{y}_{d,t} = \sum_{m=1}^M \mathbf{f}_t^H \mathbf{h}_{m,t} a_{m,t}\frac{g_{m,t}[d]}{v_{m,t}}+\mathbf{f}_t^H \mathbf{n}_{d,t}.
    \end{equation}

The server uses $\mathbf{\hat{r}}_t \triangleq [\hat{r}_t[1],...,\hat{r}_t[D]]^T$ to update the current global model $\mathbf{w}_t$ based on \eqref{ModelUpdate}. The FL system with over-the-air aggregation is shown in Fig. \ref{systemmodel}.

\subsection{Problem Formulation}

Aiming for a power-efficient communication design for FL, our goal is to minimize the total uplink transmit power over all devices during each communication round while ensuring the convergence of the model training to the optimal model. Specifically, we optimize the transmit weights $a_{m,t}$'s of all devices and the receive beamforming vector $\mathbf{f}_t$ in each communication round $t$ to minimize the sum transmit power under the global model $\mathbf{w}_t$ convergence guarantee. This leads to a joint transmit and receive beamforming optimization problem in each communication round $t$ that is formulated as follows:
\begin{subequations}\label{P1}
    \begin{align}
\min_{\mathbf{f}_t, \{a_{m,t}\} } \quad & \sum_{m=1}^M \left|a_{m,t}\right|^2 \label{P1:objective}\\
    \textrm{s.t.} \quad &
    \underset{\tau \to \infty}{\lim} \mathbb{E}[F(\mathbf{w}_\tau)-F(\mathbf{w^\star})] = 0, \label{P1:C1}\\
    & \left|a_{m,t}\right|^2 \le P_0, \: \forall m, \label{P1:C2}
    \end{align}
\end{subequations}
where \eqref{P1:C1} ensures that the global loss function converges to its minimum value 
$F(\mathbf{w^\star})$, as the number of communication rounds 
$\tau$ approaches infinity.

\begin{remark} We point out that even though we do not consider device selection explicitly in the problem formulation, problem \eqref{P1} captures device selection implicitly using transmit weights $a_{m,t}$ since if $a_{m,t}=0$, device $m$ will not transmit its local gradient to the server in round $t$.
\end{remark}

\section{Joint Transmit and Receive Beamforming Design}\label{BeamformingDesign}
In problem \eqref{P1}, constraint \eqref{P1:C1} on the training model convergence is difficult to handle directly. The global loss function $F(\mathbf{w})$ is unknown at the server, and its convergence criterion needs to be examined. Thus, we must first analyze the convergence of the gradient descent method in the presence of errors and beamforming. 

\subsection{Training Convergence under Communication Error}
We rewrite the global model update at the server in \eqref{ModelUpdate} as 
     \begin{align}\label{eq12}
         \mathbf{w}_{t+1} = \mathbf{w}_t - \gamma_t \:\mathbf{s}_t,
     \end{align}
where $\mathbf{s}_t \triangleq \frac{\mathcal{R}({\mathbf{\hat r}_t})}{K}$. We can equivalently rewrite $\mathbf{s}_t$ as follows:
\begin{align}\label{UpdateDef}
    \mathbf{s}_t = \nabla F(\mathbf{w}_t)+ \mathbf{e}_t,
\end{align}
where $\nabla F(\mathbf{w}_t)$ is the gradient of the global loss function at $\mathbf{w}_t$, and $\mathbf{e}_t \in \mathbb{R}^D$ is the error vector. The error $\mathbf{e}_t$ can be written as
\begin{align}\label{ErrorDef}
\mathbf{e}_t & = \mathbf{s}_t- \nabla F(\mathbf{w}_t) \nonumber\\&
\overset{\mathrm{(a)}}{=} \frac{\mathcal{R}({\mathbf{\hat r}_t})}{K}- \frac{\sum_{m=1}^M K_m \mathbf{g}_{m,t}}{K} \nonumber\\& \overset{\mathrm{(b)}}{=} \frac{1}{K} \sum_{m=1}^M\big( \frac{\mathcal{R}[\mathbf{f}^{H}_t  \mathbf{h}_{m, t} a_{m,t}]}{v_{m,t}}- K_m\big)\mathbf{g}_{m,t} + \frac{\tilde{\mathbf{n}}_t}{K},
\end{align}
where $\tilde{\mathbf{n}}_t \triangleq \mathcal{R}([\mathbf{f}_t^H \mathbf{n}_{1,t}, \ldots,\mathbf{f}_t^H \mathbf{n}_{D,t}]^T)$ and (a) is due to $\nabla F(\mathbf{w}_t) = \sum_{m=1}^M \frac{K_m}{K} \mathbf{g}_{m,t}$ based on \eqref{GlobalLossDef}, and (b) follows the definition of $\mathbf{\hat r}_t$ in \eqref{RHatDef}.

We consider the following assumptions on the global loss function, which are common in the literature on convergence analysis of FL \cite{friedlander2012hybrid, PolyakBook}.
\begin{enumerate}
\item[{\bf A1:}] $F(\mathbf{w})$ is differentiable.
\item[{\bf A2:}] $F(\mathbf{w})$ is $\mu$-strongly convex, i.e., $\exists \mu > 0$, such that
\begin{align}\label{StrongConvexity}
        F(\mathbf{w})\geq  F(\mathbf{w}^\prime) & + (\mathbf{w}-\mathbf{w}^\prime)^{T}  \nabla F(\mathbf{w}^\prime) \nonumber \\ & +\frac{\mu}{2}\|\mathbf{w}-\mathbf{w}^\prime\|^{2}_{2},  \forall \mathbf{w}, \mathbf{w}^\prime \in \mathbb{R}^D.
    \end{align}
\item[{\bf A3:}] $\nabla F(\mathbf{w})$ is $L$-Lipschitz continuous, i.e., 
$\exists L > 0$, such that
\begin{align}
\hspace{-1 em}
    \| \nabla F(\mathbf{w}) - \nabla F(\mathbf{w}^\prime)\|  \le  L \| \mathbf{w}-\mathbf{w}^\prime \|, \forall \mathbf{w}, \mathbf{w}^\prime  \in \mathbb{R}^D.
\end{align}
\end{enumerate}

In the following theorem, we establish sufficient conditions on the error $\mathbf{e}_t$ and the learning rate that ensure the convergence of the global loss function to its minimum value $F(\mathbf{w}^\star)$.

\begin{theorem}\label{Theorem1}
Under \textbf{A1}-\textbf{A3}, assume the following conditions on the error vector $\mathbf{e}_t$ and the learning rate $\gamma_t$ hold:
\begin{enumerate}
\item[{\bf C1:}] $\| \mathbb{E}[\:\mathbf{e}_t| \mathbf{w}_t\:] \| \le \alpha \| \nabla F(\mathbf{w}_t) \|$, for some $ \alpha \in [0,1), \forall t.$
\item[{\bf C2:}] $\mathbb{E}[\:\|\mathbf{e}_t\|^2| \mathbf{w}_t\:] \le \delta\| \nabla F(\mathbf{w}_t) \|^2+\beta $, for some $ \delta, \beta \ge 0, \forall t.$
 \item[{\bf C3:}] $\gamma_t \geq 0, \underset{t \to \infty}{\lim} \gamma_{t}  = 0$, $\Sigma_{t=0}^{\infty} \gamma_{t} = \infty$,
\end{enumerate}
where the expectation $\mathbb{E}[\cdot]$ in {\bf C1} and {\bf C2} is w.r.t. the receiver noise.
Then, for any initial point $\mathbf{w}_0$, the expected optimality gap between the global loss function $F(\mathbf{w}_t)$ and its minimum value  $F(\mathbf{w^\star})$ converges to zero, i.e., $\underset{t \to \infty}{\lim} \mathbb{E}[F(\mathbf{w}_t)-F(\mathbf{w^\star})] = 0$.
\begin{proof}
See Appendix \ref{Appendix:A}. 
\end{proof}

\end{theorem}

Conditions \textbf{C1} and \textbf{C2} in Theorem~\ref{Theorem1} limit the first and second-order statistics of the error $\mathbf{e}_t$, in terms of $\| \nabla F(\mathbf{w}_t)\|$, respectively. 
Condition \textbf{C3} requires the learning rate to diminish over $t$ while remaining non-summable.

\begin{remark} One way to satisfy \textbf{C3} is by choosing a constant learning rate during the first $T$ communication rounds, i.e., ($\gamma_t = \gamma, 0 \le t \le T$) and setting $\gamma_t $ as a harmonic series for the rest of the rounds, i.e., ($\gamma_t = \frac{\gamma}{t}, t > T$).
\end{remark}

\subsection{Problem Reformulation }
\label{sec:prob reformulation}

Based on Theorem~\ref{Theorem1}, if the learning rate during the training process meets \textbf{C3}, then to achieve convergence to the optimum in \eqref{P1:C1}, it is sufficient to satisfy \textbf{C1} and \textbf{C2}.
Thus, to make the per-round power minimization problem \eqref{P1} more tractable, we replace constraint \eqref{P1:C1} with \textbf{C1} and \textbf{C2} and reformulate problem \eqref{P1} as follows:
\begin{subequations}\label{P2}
    \begin{align}
    \min_{ \mathbf{f}_t, \{a_{m,t}\} } &  \sum_{m=1}^M \left|a_{m,t}\right|^2 \\
    \textrm{s.t.} \quad &
    \big\| \mathbb{E}\big[\mathbf{e}_t| \mathbf{w}_t\big] \big\| \le \alpha \| \nabla F(\mathbf{w}_t) \|, \label{P2:C1}\\
    & \mathbb{E}\big[\|\mathbf{e}_t\|^2| \mathbf{w}_t\big] \le \delta\| \nabla F(\mathbf{w}_t) \|^2+\beta, \label{P2:C2}\\
    & \left|a_{m,t}\right|^2 \le P_0, \: \forall m. \label{P2:C3}
    \end{align}
\end{subequations}

Based on Theorem~\ref{Theorem1}, a solution to problem \eqref{P2} is always feasible to the original problem (7), provided C3 is satisfied. However, the reformulated problem \eqref{P2} still cannot be solved. In particular, in communication round $t$, the server aims to solve problem \eqref{P2} to determine the receive beamforming and transmit weights of all devices to be used in this round. However, from \ref{ErrorDef}, the left-hand side (LHS) of \eqref{P2:C1} and \eqref{P2:C2} are functions of local gradients $\{\mathbf{g}_{m,t}\}$, which are unknown to the server. Furthermore, $\| \nabla F(\mathbf{w}_t) \|$ on the right-hand side (RHS) of \eqref{P2:C1} and \eqref{P2:C2} is unknown to the server as the server has no information about the global gradient.
To overcome these challenges, we first develop upper bounds to the LHS of \eqref{P2:C1} and \eqref{P2:C2} and use them in \eqref{P2:C1} and \eqref{P2:C2} instead. The following two propositions provide these two upper bounds.

\begin{proposition}\label{Proposition1} Given $\mathbf{w}_t$, $\big\| \mathbb{E}\big[\mathbf{e}_t| \mathbf{w}_t\big] \big\|$ is upper bounded by
\begin{align} \label{error_bound}
\big\| \mathbb{E}\big[\mathbf{e}_t| \mathbf{w}_t\big] \big\|  \le  \frac{\sqrt{D}}{K} \sum_{m=1}^M  \big|K_m v_{m,t}-\mathcal{R}[\mathbf{f}_t^H  \mathbf{h}_{m, t} a_{m,t}] \big|.
\end{align}
\end{proposition}
\IEEEproof
See Appendix \ref{Appendix:B}. 
\endIEEEproof
\begin{proposition}\label{Proposition2} Given $\mathbf{w}_t$, $\mathbb{E}\big[\|\mathbf{e}_t\|^2| \mathbf{w}_t\big]$ is upper bounded by
\begin{align}\label{Proposition2:eq}
\mathbb{E}\big[&\|\mathbf{e}_t\|^2| \mathbf{w}_t\big]\nonumber \\ & \le   \frac{D}{K^2} \Big(\sum_{m=1}^M \left|K_m v_{m,t}- \mathcal{R}[\mathbf{f}_t^H \mathbf{h}_{m, t} a_{m,t}]\right| \Big)^2+ \frac{ D \sigma_n^2 \| \mathbf{f}_t\|^2}{ 2K^2}.
\end{align}
\end{proposition}
\IEEEproof
See Appendix \ref{Appendix:C}. 
\endIEEEproof
To address the second challenge on unknown $ \|\nabla F(\mathbf{w}_t)\|$ at the RHS of \eqref{P2:C1} and \eqref{P2:C2}, we can approximate it by 
\begin{align}
    \|\nabla F(\mathbf{w}_t)\| & \overset{\mathrm{(a)}}{\le} \sum_{m=1}^M \frac{K_m}{K} \|\mathbf{g}_{m,t}\| 
    \overset{\mathrm{(b)}}{=} \sum_{m=1}^M \frac{K_m}{K} \sqrt{D} v_{m,t} \label{App:3}
    \end{align}
where (a) is obtained by using $\nabla F(\mathbf{w}_t)= \sum_{m=1}^M \frac{K_m}{K} \mathbf{g}_{m,t}$ and applying the triangle inequality, and (b) follows the definition of $v_{m,t}$. 
Note that the inequality in (a) can be replaced by equality when all the local gradients $\{\mathbf{g}_{m,t}\}$ have the same direction. 
In the situations where the local gradients are similar (such as i.i.d. data distribution over devices), \eqref{App:3} provides a close approximation for $\|\nabla F(\mathbf{w}_t)\|$. We denote this approximation by $V_t \triangleq \frac{\sqrt{D}}{K} \sum_{m=1}^M K_m v_{m,t}$. 

Based on Propositions~\ref{Proposition1} and~\ref{Proposition2}, and the approximation in \eqref{App:3}, we further modify both sides of \eqref{P2:C1} and \eqref{P2:C2} and arrive at the following optimization problem in each round $t$:
\vspace*{-0.6em}
\begin{subequations}\label{P5}
    \begin{align}
\hspace*{-1em}  \min_{\mathbf{f}_t, \{a_{m,t}\}} & \sum_{m=1}^M \left|a_{m,t}\right|^2
\label{P5:objective}\\
    \textrm{s.t.} \quad &
    \frac{\sqrt{D}}{K} \sum_{m=1}^M \left|K_m v_{m,t}-\mathcal{R}[\mathbf{f}_t^H  \mathbf{h}_{m, t} a_{m,t}]\right| \le \alpha V_t, \label{P5:C1}\\
    & \frac{D}{K^2} \Big(\sum_{m=1}^M \left|K_m v_{m,t}- \mathcal{R}[\mathbf{f}_t^H \mathbf{h}_{m, t} a_{m,t}]\right| \Big)^2 \nonumber\\ & \qquad \qquad \qquad + \frac{ D \sigma_n^2 \| \mathbf{f}_t\|^2}{ 2K^2}  \le \delta V_t^2+\beta, \label{P5:C2}\\
    & \left|a_{m,t}\right|^2 \le P_0, \: \forall m. \label{P5:C3}
    \end{align}
\end{subequations}
Note that $\alpha$, $\delta$, and $\beta$ are the pre-set constant parameters to guarantee convergence by Theorem~\ref{Theorem1}. 
Setting $\alpha$ to a larger value allows a larger value of the expected error $ \| \mathbb{E}[\:\mathbf{e}_t| \mathbf{w}_t\:] \| $. 
Similarly, increasing $\delta$ and $\beta$ leads to a higher value of $\mathbb{E}[\|\mathbf{e}_t\|^2| \mathbf{w}_t]$. Therefore, increasing these constants results in a larger deviation of the model update from the global gradient, which reduces the convergence rate and degrades the FL learning performance. On the other hand, although smaller values of $\alpha$, $\delta$, and $\beta$ reduce the error of the model update and improve learning performance, they lead to larger power consumption at devices due to a smaller feasible set of the optimization problem.
Thus, the values of these parameters affect the trade-off between the learning performance and the device power consumption.

\begin{remark}
We point out that for the approximation used at the RHS of constraints \eqref{P5:C1} and \eqref{P5:C2}, if $V_t$ satisfies $ \frac{V_t}{\|\nabla F(\mathbf{w}_t)\|} < \frac{1}{\alpha}, \forall t$, there exists a set of values $\alpha^{\prime} \ge \alpha$, $\beta^{\prime} = \beta$, and $\delta^{\prime} \ge \delta$ for which \textbf{C1} and \textbf{C2} in Theorem~\ref{Theorem1} are satisfied. In this case, FL convergence to the optimum is ensured by any feasibility solution to problem \eqref{P5}. 
Therefore, in scenarios where $\| \nabla F(\mathbf{w}_t)\|$ may significantly differ from $V_t$ (e.g., non-i.i.d. data distribution across devices), selecting smaller value of $\alpha$ for problem \eqref{P5} can still guarantee FL training convergence.
\end{remark}

\begin{remark}
 In problem \eqref{P5}, it is possible to assign significantly large values to $\delta$ and $\beta$ for constraint \eqref{P5:C2} and achieve near-zero power $\left|a_{m,t}\right|$, by choosing $\|\mathbf{f}_t\|$ to be very large. While this choice of solution can significantly reduce the power consumption per iteration, the convergence to the optimum is only assured under Assumptions (\textbf{A1}-\textbf{A3}). However, although these assumptions are necessary for tractable convergence analysis, they do not hold in many machine learning applications. Thus, choosing arbitrarily large values for $\delta$, $\beta$, and $\alpha$, may significantly reduce the robustness in ensuring convergence towards an optimal model for the general loss functions. Consequently, in our experiments where the loss function may not satisfy \textbf{A1}-\textbf{A3}, we
select $\delta$, $\beta$, and $\alpha$ such that we achieve the same convergence rate as the benchmark methods ensuring a fair comparison of power consumption among different methods.
\end{remark}

\subsection{Proposed Joint Beamforming Algorithm} \label{PoFL}
 Problem \eqref{P5} is non-convex since the constraints  \eqref{P5:C1} and \eqref{P5:C2} are not jointly convex in $\mathbf{f}_t$ and $\{a_{m,t}\}$. However, the problem is bi-convex, as all constraints are convex in $\{a_{m,t}\}$, as well as in $\mathbf{f}_t$. Thus, we can apply the alternating optimization approach and solve problem   \eqref{P5} alternatingly with respect to (w.r.t.)  $\{a_{m,t}\}$ and  $\mathbf{f}_t$. This iterative approach is guaranteed to find a partial optimum solution to  problem   \eqref{P5}  \cite{BiConvex-Theorem4.7}. Below, we discuss each optimization subproblem.

 \subsubsection{\textbf{Transmit Weights Optimization}}\label{optimization-step2} 
Given the receive beamforming vector $\mathbf{f}_t$, we optimize the transmit weights $\{a_{m,t}\}$ in problem \eqref{P5}.
Observing that  the LHS of constraints \eqref{P5:C1} and \eqref{P5:C2} contain the same summation term, we  combine \eqref{P5:C1} and \eqref{P5:C2}  into a single constraint and transform problem  \eqref{P5} into the following equivalent form:
\begin{subequations}\label{P7}
    \begin{align}
    \min_{\{a_{m,t}\}} \quad & \sum_{m=1}^M \left|a_{m,t}\right|^2\\
    \textrm{s.t.} \quad &
    \frac{\sqrt{D}}{K} \sum_{m=1}^M \big|K_m v_{m,t}-\mathcal{R}[\mathbf{f}_t^H  \mathbf{h}_{m, t} a_{m,t}]\big| \le  R_t,
    \label{P7:C1}\\
    & \left|a_{m,t}\right|^2 \le P_0, \: \forall m, \label{P7:C2}
    \end{align}
\end{subequations}
where $R_t \triangleq  \min \big(\alpha V_t, \sqrt{\delta V_t^2 +\beta - \frac{D \sigma_{n}^{2} \|\mathbf{f}_{t} \|^{2}}{2K^{2}}}\big)$. We show in the following proposition that we can further simplify problem  \eqref{P7} by only considering the real values for $\{a_{m,t}\}$.

\begin{proposition}\label{Proposition3} 
For the optimal solution to problem \eqref{P7}, denoted by $\{a_{m,t}^\star\}$, its phase $\angle a_{m,t}^\star \in [0, 2\pi)$ satisfies
\begin{align}\label{opt_phase}
\angle a_{m,t}^\star = - \angle \mathbf{f}_t^H \mathbf{h}_{m, t}, \ \forall m.
\end{align}
\end{proposition}
\IEEEproof
See Appendix \ref{Appendix:D}. 
\endIEEEproof
Since the phase of the optimal  $a_{m,t}^\star$ to problem \eqref{P7} can be obtained from \eqref{opt_phase}, we only need to find $|a_{m,t}|$, $\forall m$. Define $b_{m,t} \triangleq |a_{m,t}|$. We rewrite problem \eqref{P7}  w.r.t. $\{ b_{m,t}\}$ as 
\begin{subequations}\label{P10}
    \begin{align}
    \min_{\{b_{m,t}\}} \quad & \sum_{m=1}^M b_{m,t}^2\\
    \textrm{s.t.} \quad &
    \frac{\sqrt{D}}{K} \sum_{m=1}^M \left|K_m v_{m,t}-|\mathbf{f}_t^H  \mathbf{h}_{m, t}|\: b_{m,t}\right| \le R_t, \label{P10:C1}\\
    & 0 \le b_{m,t} \le \sqrt{P_0}, \: \forall m. \label{P10:C2}
    \end{align}
\end{subequations}
We further convert constraint \eqref{P10:C1} into a more amenable form without $|\cdot|$ operation. In particular, the following proposition shows that problem  \eqref{P10} is equivalent to a QP problem.

\begin{proposition}\label{Proposition4} 
Problem \eqref{P10} is equivalent to the following QP problem:
\begin{subequations}\label{P11}
    \begin{align}
    \min_{\{b_{m,t}\}} \quad & \sum_{m=1}^M b_{m,t}^2\\
    \textrm{s.t.} \quad &
    \frac{\sqrt{D}}{K} \sum_{m=1}^M\big( K_m v_{m,t}-|\mathbf{f}_t^H  \mathbf{h}_{m, t}|\: b_{m,t}\big) \le R_t,
    \label{P11:C1} \\
    & 0 \le b_{m,t} \le \min\left\{ \sqrt{P_0}, \  \frac{K_m v_{m,t}}{|\mathbf{f}_t^H \mathbf{h}_{m, t}|}\right\}, \: \forall m. \label{P11:C3}
    \end{align}
\end{subequations}
\end{proposition}
\IEEEproof
See Appendix \ref{Appendix:E}. 
\endIEEEproof

Since problem \eqref{P11} is a QP problem, it can be efficiently solved using standard QP solvers such as CVXOPT \cite{Grant&Boyd:CVX}.

\subsubsection{\textbf{Receive Beamforming Optimization}}\label{AlternatingApproach-step2} 

Note that the receive beamforming vector $\mathbf{f}_t$ only affects the constraints but not the objective function.
Thus,
the goal for optimizing  $\mathbf{f}_t$ is to enlarge the feasible set of $\{a_{m,t}\}$ for transmit weight optimization problem of the subsequent iteration. This will ensure that the objective value of problem \eqref{P7} is monotonically decreasing over iterations. 

Based on the above discussion, given $\{a_{m,t}\}$, we minimize the LHS of constraint \eqref{P5:C2} w.r.t. $\mathbf{f}_t$ under constraint    \eqref{P5:C1} as follows:
\begin{subequations}\label{P8}
    \begin{align}
    \min_{\mathbf{f}_t} \quad & \frac{D}{K^2} \big(\!\sum_{m=1}^M  \left|K_m v_{m,t}- \mathcal{R}[\mathbf{f}_t^H \mathbf{h}_{m, t} a_{m,t}]\right| \big)^2 \! \!+ \!\frac{ D \sigma_n^2 \| \mathbf{f}_t\|^2}{ 2K^2}\\
    \textrm{s.t.} \quad &
    \frac{\sqrt{D}}{K} \sum_{m=1}^M  \left|K_m v_{m,t}-\mathcal{R}[\mathbf{f}_t^H  \mathbf{h}_{m, t} a_{m,t}]\right|  \le \alpha V_t. \label{P8:C1}
    \end{align}  
\end{subequations}

 Note that the LHS of constraint  \eqref{P5:C2} represents the expected deviation from the true gradient. Minimizing it w.r.t. $\mathbf{f}_t$ increases the gap between the LHS and the RHS of the constraint, leading to a larger feasible set of  $\{a_{m,t}\}$ for the subproblem of transmit weights optimization  in problem   \eqref{P7} in the subsequent iteration. 

The expressions  of objective and constraint functions  in problem \eqref{P8} contains $|\cdot|$ operation. To simplify it, we introduce $M$ real-valued auxiliary  variables forming a vector as $\mathbf{r}_t \triangleq [r_{1,t}, r_{2,t}, ..., r_{M,t}]^T$, and transform  problem \eqref{P8} into the following equivalent problem w.r.t. $\mathbf{f}_t$ and $\mathbf{r}_t$:
\begin{subequations}\label{P9}
    \begin{align}
    \min_{\mathbf{f}_t, \mathbf{r}_t} \quad & | \mathbf{1}^T\mathbf{r}_t|^2 + \frac{ \sigma_n^2 }{ 2} \| \mathbf{f}_t\|^2\\
    \textrm{s.t.} \quad &
    K_m v_{m,t}-\mathcal{R}[\mathbf{f}_t^H  \mathbf{h}_{m, t} a_{m,t}] \le r_{m,t}, \forall m, \label{P9:C2}\\
    \quad &
    \mathcal{R}[\mathbf{f}_t^H  \mathbf{h}_{m, t} a_{m,t}]- K_m v_{m,t} \le r_{m,t},
    \forall m, \label{P9:C3}\\  \quad &
    \frac{\sqrt{D}}{K}\mathbf{1}^T\mathbf{r}_t  \le \alpha V_t,
    \end{align}
    \end{subequations}
where $\mathbf{1}$ is an $M\times 1$ all-one vector. We see that problem \eqref{P9} is a QP problem, and thus, we can solve it using standard QP solvers such as CVXOPT  to obtain the solution efficiently.

\subsubsection{\textbf{Final Scaling of Receive Beamforming and Transmit Weights}}\label{optimization-step3} 
The two subproblems   \eqref{P7} and  \eqref{P8} are solved alternatingly until convergence.  Note that  under the solution    ($\mathbf{f}^\text{\tiny AO}_t,\{a^\text{\tiny AO}_{m,t}\})$ obtained via the alternating optimization approach, constraint  \eqref{P5:C2} may not hold with equality. In this case, we can further reduce the objective value  by scaling $\mathbf{f}^\text{\tiny AO}_t$  and $\{a^\text{\tiny AO}_{m,t}\}$ such that constraint \eqref{P5:C2} is satisfied with equality. Specifically,  we scale $\mathbf{f}^\text{\tiny AO}_t$  and $\{a^\text{\tiny AO}_{m,t}\}$ as follows:
\begin{align}\label{scaling}
    \mathbf{f}_t &= p_t \: \mathbf{f}^\text{\tiny AO}_t, \quad a_{m,t} = \frac{a^\text{\tiny AO}_{m,t}}{p_t}, \ \forall m. 
\end{align}
where  $p_t \in \mathbb{R}^+$ is the scaling factor.

We note that the first term at the LHS of   \eqref{P5:C2} depends only on the product of     $\mathbf{f}_t$  and $\{a_{m,t}\}$ and thus will remain unchanged after the scaling using \eqref{scaling}. Consequently, to change constraint \eqref{P5:C2}  from inactive
to  active, we have $p_t>1$. In other words, $\|\mathbf{f}_t\|$ is scaled up, and this allows us to proportionally scale down $\{a_{m,t}\}$, leading to reduced device power consumption.

To obtain the value of $p_t$,
we note that at the optimality of problem \eqref{P11},  constraint \eqref{P11:C1} is always  satisfied with equality. Otherwise, we can further decrease $\{b_{m,t}\}$  to increase the value of the LHS of \eqref{P11:C1}, leading to a lower objective value. Recall that $R_t= \min \{\alpha V_t,\sqrt{\delta V_t^2+\beta - \frac{D\sigma_n^2 \|\mathbf{f}_t\|^2}{2K^2}}\}$. Its value indicates which of  constraints \eqref{P5:C1} and  \eqref{P5:C2} holds with equality. If constraint \eqref{P5:C2} is inactive, we scale it using   \eqref{scaling}, such that both constraints \eqref{P5:C1} and  \eqref{P5:C2} hold with equality. Thus, the scaling factor $p_t$ is given by
\begin{align}
\label{scalingfactor}
p_t= \frac{1}{\| \mathbf{f}^\text{\tiny AO}_t\|}\frac{\sqrt{2}K}{\sqrt{D}\sigma_n}\sqrt{(\delta-\alpha^2) V_t^2 + \beta}.
\end{align}

\subsection{{Proposed Initialization Method}}\label{Init} 
Our proposed algorithm above using alternating optimization to solve problem \eqref{P5} requires a feasible receive beamforming vector  $\mathbf{f}_t$ as the initial point. In other words, we need to find $\mathbf{f}_t$, such that there exist at least one set of $\{a_{m,t}\}$ that satisfy constraints \eqref{P5:C1}--\eqref{P5:C3}.
  Furthermore, a good initial point is desirable to accelerates  convergence. In this subsection, we propose an initialization method to determine a feasible initial point $\mathbf{f}_t$.

For any given receive beamforming vector $\mathbf{f}_t$, we can always find $\{a_{m,t}\}$ to make the summation terms at the 
LHS of constraints \eqref{P5:C1} and \eqref{P5:C2}  zero, where $a_{m,t}$ is given by 
\begin{align}\label{Zeroforcing}
    a_{m,t}= \frac{K_m v_{m,t}}{\mathbf{f}_t^H  \mathbf{h}_{m, t}}, \quad \forall m.
\end{align}
Thus, there always exists a set of $\{a_{m,t}\}$ satisfying constraint \eqref{P5:C1} for any given $\mathbf{f}_t$. In fact, recall from Proposition~\ref{Proposition1} that the LHS of \eqref{P5:C1} is the upper bound on $\|\mathbb{E}[\mathbf{e}_t|\mathbf{w}_t]\|$. Thus, setting $a_{m,t}$ as in \eqref{Zeroforcing} leads to $\mathbb{E}[\mathbf{e}_t|\mathbf{w}_t]= \mathbf{0}$, i.e., zero-mean error, and hence, $\mathbf{s}_t$ in this case becomes an unbiased estimate of $\nabla F(\mathbf{w}_t)$.

Next, to expand the feasible set of  $\{a_{m,t}\}$ for problem \eqref{P5}, we intend to find $\mathbf{f}_t$ that minimizes the LHS of constraint \eqref{P5:C2} while satisfying constraint \eqref{P5:C3}. Following \eqref{Zeroforcing}, since the first term in constraint \eqref{P5:C2} becomes zero, we have the equivalent optimization problem w.r.t. $\mathbf{f}_t$ as follows:
\begin{subequations}\label{feasiblityProb}
\begin{align}
\min_{\mathbf{f}_t} \quad &  \frac{ D \sigma_n^2 \| \mathbf{f}_t\|^2}{ 2K^2} \label{feasiblityProb_a} \\
\text{s.t.}  \quad &  |\mathbf{f}_t^{H} \mathbf{h}_{m, t}| ^2  \ge \frac{K_m^2 v_{m,t}^2}{P_0},  \quad \forall m.
\end{align}
\end{subequations}

Note that the beamforming problem \eqref{feasiblityProb} is equivalent to the classical single-group multicast beamforming quality-of-service (QoS) problem \cite{Sidiropoulos&etal:TSP2006, dong2020multi}. In downlink multicasting, the base station (BS)  uses a multicast beamforming vector to simultaneously transmit a common message to all devices. The QoS problem is to optimize the  the multicast beamformer to minimize the BS transmit power while meeting the SNR target at each device. We see that under over-the-air aggregation, the uplink receive beamforming  problem  \eqref{feasiblityProb} is also a power minimization problem subject to the minimum receive SNR\ target of each device. Specifically, $\mathbf{f}_t$ represents the transmit beamformer, and $\frac{K_m^2 v_{m,t}^2}{P_0}$ denotes the SNR target for device $m$. Although the multicast beamforming problem is an NP-hard, its optimal structure is known  \cite{dong2020multi}, and the successive convex approximation (SCA) method can be used to compute the beamforming solution, which offers a convergence guarantee to a stationary point of problem  \eqref{feasiblityProb}. 

Using the solution $\mathbf{f}_t$  from problem \eqref{feasiblityProb}, we evaluate the objective value in \eqref{feasiblityProb_a}, denoted by $q_t$. There are two cases: 1) If $q_t \le \delta V_t^2 + \beta$: This means our approach above has obtained a feasible point $\{a_{m,t}\}$ that satisfies all the constraints \eqref{P5:C1} -- \eqref{P5:C3}; 2) If $q_t>\delta V_t^2 + \beta$: The above approach fails to find a feasible point $\{a_{m,t}\}$ for problem \eqref{P5}.  Even if problem \eqref{P5} is feasible, the second case may still occur since the above approach of using the relation in \eqref{Zeroforcing} is suboptimal. In this case, we choose to generate $\mathbf{f}_t$ randomly until it is feasible. 
 
We point out that the solution to problem \eqref{feasiblityProb} is typically feasible to problem \eqref{P5} in our experiments, 
so long the values of $\alpha$ and $\beta$ for constraints \eqref{P5:C1} and \eqref{P5:C2} are not chosen to be too small, causing the feasible set for the problem to be too small.

\subsection{Convergence and Complexity Analysis}

Since we solve the two subproblems  \eqref{P7} and \eqref{P8} optimally in  the alternating optimization, the sum transmit power objective value in problem   \eqref{P5} is non-increasing over  iterations. Since it is also bounded from zero, the iterative procedure  is guaranteed to converge to a partial optimum\cite{BiConvex-Theorem4.7}. The final scaling step in Section~\ref{optimization-step3}   provides an improved solution by further reducing the objective value without violating the constraints. 

The main computational complexity of the proposed algorithm is at solving QP problems   \eqref{P11} and \eqref{P9}. Problem \eqref{P11}  contains $M$ variables and $2M+1$ constraints. The computational complexity to solve it via standard convex solvers is  $\mathcal{O}(M^3)$.
Problem  \eqref{P9} has $M+N$ variables and the computational complexity  is  $\mathcal{O}((M+N)^3)$.
For the initialization method,  problem~\eqref{feasiblityProb} has  $N$ variables and $M$ constraints. It  can be solved by exploring the optimal structure with the SCA method as discussed earlier, and the computational complexity is    $\mathcal{O}(I_{\text{max}}\text{min}(M, N)^3 )$ with standard convex solvers, where $I_{\text{max}}$ is the maximum number of the SCA iterations. 
Thus, the overall computational complexity of our proposed algorithm with the initialization method for each communication round is $\mathcal{O}((M+N)^3)$.

\section{CSI-Error-Aware Joint Beamforming Design}\label{impsection}
The proposed algorithm in Section \ref{BeamformingDesign} for the join beamforming design assumes the perfect CSI knowledge at the server. In practice,  only the estimated CSI can be obtained at the server. In this case, directly apply the algorithm in Section \ref{BeamformingDesign} by substituting the actual channels with their estimates may no longer guarantee the FL convergence, since constraints \eqref{P5:C1} and \eqref{P5:C2} under the actual channels are not guaranteed to be satisfied. In this section, we aim to extend our method to accommodate imperfect CSI and design joint beamforming, taking into account the channel estimation error at the server.

Let  $\hat{\mathbf{h}}_{m,t}$ denote the estimate of the uplink channel $\mathbf{h}_{m,t}$ between  device $m$ and the server at the beginning of round $t$, which is given by $
   \hat{\mathbf{h}}_{m,t} = \mathbf{h}_{m,t}+ \tilde{\mathbf{h}}_{m,t}, $ where $\tilde{\mathbf{h}}_{m,t}$ is the  channel estimation error. We assume both $\mathbf{h}_{m,t}$ and 
$\tilde{\mathbf{h}}_{m,t}$ follow the complex Gaussian distribution, i.e., $\mathbf{h}_{m,t} \sim \mathcal{CN}(\mathbf{0},\sigma_{h_m}^2 \mathbf{I})$, and $\tilde{\mathbf{h}}_{m,t} \sim \mathcal{CN}(\mathbf{0}, \epsilon \sigma_{h_m}^2 \mathbf{I})$, where $\sigma_{h_m}^2$ is the variance of the uplink channel and $\epsilon \sigma_{h_m}^2$ is the mean squared error of the channel estimate with $\epsilon \in [0, 1)$ indicating the normalized error. We assume the mean squared error of the channel estimate, $\epsilon \sigma_{h_m}^2$, is available at the server.

The error $\mathbf{e}_t$  in \eqref{ErrorDef} based on local gradients can be rewritten as
\begin{align}\label{ErrorDef-ICSI}
\hspace*{-.5em}    \mathbf{e}_t = \!\frac{1}{K}\! \sum \limits_{m}\!\Big( \frac{\mathcal{R}[\mathbf{f}^{H}_t  (\hat{\mathbf{h}}_{m,t} \!-\!\tilde{\mathbf{h}}_{m,t})  a_{m,t}]}{v_{m,t}}- K_m\!\Big)\mathbf{g}_{m,t} \!+ \!\frac{\tilde{\mathbf{n}}_t}{K}.
\end{align}
Given  $\mathbf{w}_t$, we extend the bounds on the first and second-order statistics of  $\mathbf{e}_t$ in  Propositions~\ref{Proposition1} and \ref{Proposition2} to the general case with the knowledge of  estimated CSI $\hat{\mathbf{h}}_{m,t}$'s.
\begin{proposition}\label{Proposition1-imperfect}
Given $\mathbf{w}_t$ and $\hat{\mathbf{h}}_{m, t}$, $\forall m$, the error bound in \eqref{error_bound} holds by replacing $\mathbf{h}_{m, t}$ with $\hat{\mathbf{h}}_{m, t}$, $\forall m$. 
\end{proposition}
\IEEEproof
The proof straightforwardly follows the proof of Proposition 1.
\endIEEEproof

\begin{proposition}\label{Proposition2-imperfect} Given $\mathbf{w}_t$ and $\hat{\mathbf{h}}_{m, t}$, $\forall m$, $\mathbb{E}[\|\mathbf{e}_t\|^2| \mathbf{w}_t]$ is upper bounded by
        \begin{align}
\hspace*{-.7em}            \mathbb{E}\big[\big\|\mathbf{e}_t\big\|^2\big| \mathbf{w}_t\big]  & \le \frac{D}{K^2} \Big(\sum_{m=1}^M \big|K_m v_{m,t}- \mathcal{R}[\mathbf{f}_t^H \hat{\mathbf{h}}_{m, t} a_{m,t}] \big| \Big)^2 \nonumber \\ 
& \   + \frac{ D \sigma_n^2 \| \mathbf{f}_t\|^2}{ 2K^2} + \frac{\epsilon D   \| \mathbf{f}_t\|^2}{ 2K^2} \sum_{m=1}^M \sigma_{h_m}^2 \left|a_{m,t}\right|^2.
        \end{align}
\end{proposition}
\IEEEproof
See Appendix \ref{Appendix:G}. 
\endIEEEproof

We note that Propositions \ref{Proposition1} and  \ref{Proposition2} are the special  cases of Propositions \ref{Proposition1-imperfect} and \ref{Proposition2-imperfect}, respectively, for $\epsilon = 0$.

Similar to the reformulation of problem \eqref{P2} to problem \eqref{P5} under perfect CSI, given  estimated CSI  $\hat{\mathbf{h}}_{m, t}$'s, and based on the bounds in Propositions \ref{Proposition1-imperfect} and \ref{Proposition2-imperfect}, we  reformulate  problem \eqref{P2}  to the following problem:
\begin{subequations}\label{ImperfectCSIProblem}
    \begin{align}
\hspace*{-1em}  \min_{\mathbf{f}_t, \{a_{m,t}\}} & \sum_{m=1}^M \left|a_{m,t}\right|^2 \\
    \textrm{s.t.} \quad &
    \frac{\sqrt{D}}{K} \sum_{m=1}^M \big|K_m v_{m,t}-\mathcal{R}[\mathbf{f}_t^H  \hat{\mathbf{h}}_{m, t} a_{m,t}]\big| \le \alpha V_t, \label{ImperfectCSIProblem:c1} \\
    & \frac{D}{K^2} \Big(\sum_{m=1}^M \big|K_m v_{m,t}- \mathcal{R}[\mathbf{f}_t^H \hat{\mathbf{h}}_{m, t} a_{m,t}]\big| \Big)^2 \nonumber\\ &   + \frac{ D \| \mathbf{f}_t\|^2}{ 2K^2}\big(\sigma_n^2+ \epsilon\!\sum_{m=1}^M \!\sigma_{h_m}^2 \left|a_{m,t}\right|^2\big) \! \le \! \delta V_t^2+\beta, \label{ImperfectCSIProblem:c2}\\
    & \left|a_{m,t}\right|^2 \le P_0, \: \forall m. \label{ImperfectCSIProblem:c3}
    \end{align}
\end{subequations}

Compared with problem  \eqref{P5}, only constraint \eqref{ImperfectCSIProblem:c2} takes a  form different from constraint \eqref{P5:C2}, where an extra term due to channel estimation error appears at the LHS of  \eqref{ImperfectCSIProblem:c2} as a function of  $\epsilon$ and $\sigma_{h_m}^2$. Nonetheless, problem \eqref{ImperfectCSIProblem} is still bi-convex in $\mathbf{f}_t$ and $\{a_{m,t}\}$, and we can use the  procedure of the proposed algorithm in Sections~\ref{optimization-step2}--~\ref{optimization-step3}: Alternatingly solving the transmit weight optimization and the receive beamforming optimization subproblems, followed by the final scaling step. However,  the two subproblems under estimated CSI   $\hat{\mathbf{h}}_{m, t}$'s  have different structures. Below, we focus on the solution to each subproblem.  
\subsection{Proposed Algorithm}

\subsubsection{\textbf{Transmit Weights Optimization}}\label{optimization-step1-IMCSI} 
Note that, given $\mathbf{f}_t$, different from constraint \eqref{P7:C1} in problem \eqref{P7}, we no longer can combine constraints \eqref{ImperfectCSIProblem:c1} and \eqref{ImperfectCSIProblem:c2} into a single constraint, since the second term at the LHS of  \eqref{ImperfectCSIProblem:c2} is also a function of $\{a_{m,t}\}$.  Following the similar argument in Proposition \ref{Proposition3}, it is sufficient to solve  problem \eqref{ImperfectCSIProblem} w.r.t. $|a_{m,t}|$, $\forall m$. Let  $b_{m,t} \triangleq |a_{m,t}|, \forall m$. Given $\mathbf{f}_t$, we can rewrite problem \eqref{ImperfectCSIProblem} in terms of $\{b_{m,t}\}$ as follows:
\begin{subequations}\label{ImperfectCSIProblem2}
    \begin{align}
\hspace*{-1em}    \min_{\{b_{m,t}\}}  & \sum_{m=1}^M b_{m,t}^2 \\
    \textrm{s.t.} \quad &
    \frac{\sqrt{D}}{K} \sum_{m=1}^M \big|K_m v_{m,t}-|\mathbf{f}_t^H  \hat{\mathbf{h}}_{m, t}| b_{m,t} \big| \le \alpha V_t, \label{ImperfectCSIProblem2:c1} \\
    & \frac{D}{K^2} \Big(\sum_{m=1}^M \big|K_m v_{m,t}- |\mathbf{f}_t^H \hat{\mathbf{h}}_{m, t}| b_{m,t}\big| \Big)^2 \nonumber\\ &  \quad + \frac{ D \| \mathbf{f}_t\|^2}{ 2K^2}\big(\sigma_n^2+ \epsilon\sum \limits_{m=1}^M \sigma_{h_m}^2 b_{m,t}^2\big) \le \delta V_t^2+\beta, \label{ImperfectCSIProblem2:c2}\\
    & 0 \le b_{m,t} \le \sqrt{P_0}, \: \forall m.
    \end{align}
\end{subequations}

Problem \eqref{ImperfectCSIProblem2} cannot be readily solved due to the terms with $|\cdot|$ operation in constraints \eqref{ImperfectCSIProblem2:c1} and \eqref{ImperfectCSIProblem2:c2}.
We further transform problem \eqref{ImperfectCSIProblem2}
into an amenable form by removing  $|\cdot|$, as described below.
\begin{proposition}\label{Proposition7} 
Problem \eqref{ImperfectCSIProblem2} is equivalent to the following problem:
\begin{subequations}\label{ImperfectCSIProblem3}
    \begin{align}
\hspace*{-1em}    \min_{\{b_{m,t}\}}  & \sum_{m=1}^M b_{m,t}^2 \\
    \textrm{s.t.} \ \ &
    \frac{\sqrt{D}}{K} \sum_{m=1}^M K_m v_{m,t}-|\mathbf{f}_t^H  \hat{\mathbf{h}}_{m, t}| b_{m,t}  \le \alpha V_t, \label{ImperfectCSIProblem3:c1} \\
    & \frac{D}{K^2} \Big(\sum_{m=1}^M K_m v_{m,t}- |\mathbf{f}_t^H \hat{\mathbf{h}}_{m, t}| b_{m,t} \Big)^2 \nonumber\\ &  \quad + \frac{ D \| \mathbf{f}_t\|^2}{ 2K^2}\big(\sigma_n^2+ \epsilon\!\sum \limits_{m=1}^M \sigma_{h_m}^2 b_{m,t}^2\big) \le \delta V_t^2+\beta, \label{ImperfectCSIProblem3:c2}
    \\& 0 \le b_{m,t} \le \min \left\{\sqrt{P_0}, \ \frac{K_m v_{m,t}}{|\mathbf{f}_t^H \hat{\mathbf{h}}_{m, t}| } \right\},\: \forall m.
    \end{align}
\end{subequations}
\end{proposition}
\begin{proof}
See Appendix \ref{Appendix:H}. 
\end{proof}

We can further express problem  \eqref{ImperfectCSIProblem3} into a compact matrix-vector form. Let $\mathbf{b}_t \triangleq [b_{1,t}, \ldots, b_{M,t}]^T$,  $ \text{\boldmath$\lambda$}_t  \triangleq  [K_1 v_{1,t}, \ldots, K_M v_{M,t}]^T$, $\mathbf{H}_t \triangleq \text{diag}(|\mathbf{f}_t^H \hat{\mathbf{h}}_{1, t}|, \ldots,|\mathbf{f}_t^H  \hat{\mathbf{h}}_{M, t}|)$, and $\boldsymbol{\Sigma} = \text{diag}(\sigma_{h_1}^2, \ldots,\sigma_{h_m}^2)$. Problem \eqref{ImperfectCSIProblem3} can be equivalently written in  $\mathbf{b}_t$ as
\begin{subequations}\label{ImperfectCSIProblem4}
    \begin{align}
    \min_{\mathbf{b}_t} \quad & \|\mathbf{b}_t\|^2\\
    \textrm{s.t.} \quad &
    \frac{\sqrt{D}}{K}\mathbf{1}^T(\text{\boldmath$\lambda$}_t - \mathbf{H}_t \mathbf{b}_t) \le \alpha V_t, \\ &
    \frac{D}{K^2} \left| \mathbf{1}^T (\text{\boldmath$\lambda$}_t - \mathbf{H}_t \mathbf{b}_t)\right|^2 +\frac{  \epsilon D \| \mathbf{f}_t\|^2}{ 2K^2} \mathbf{b}_t^T \boldsymbol{\Sigma} \mathbf{b}_t \nonumber \\ 
& \qquad \quad \quad \quad\qquad \le \delta V_t^2+\beta- \frac{ D \sigma_n^2 \| \mathbf{f}_t\|^2}{ 2K^2}, \label{QuadConst} \\
    & \mathbf{0} \preceq \mathbf{b}_t \preceq \mathbf{b}_0,  
    \end{align}
\end{subequations}
where $\mathbf{b}_0 \triangleq[b_{0,1},\ldots,b_{0,M}]^T$ with $b_{0,m}=\min \big\{\sqrt{P_0},\frac{K_m v_{m,t}}{|\mathbf{f}_t^H \hat{\mathbf{h}}_{m, t}|}\big\}$, $\forall m$. 

Note that problem  \eqref{ImperfectCSIProblem4} is a QCQP problem. We can use standard convex solvers to solve it.

\subsubsection{\textbf{Receive Beamforming Optimization}}\label{optimization-step2-IMCSI}
Given $\{a_{m,t}\}$, we follow the same approach as in problems \eqref{P8} and  \eqref{P9} to optimize $\fbf_t$. Given $\{a_{m,t}\}$,   the receive beamforming optimization problem is as in problem   \eqref{P9}, except that i) the objective function is modified to
$\left|\mathbf{1}^T\mathbf{r}_t\right|^2 + \frac{ \| \mathbf{f}_t\|^2}{ 2} \big(\sigma_n^2+ \epsilon\sum \limits_{m=1}^M \sigma_{h_m}^2 \left|a_{m,t}\right|^2\big)$, and ii) $\hbf_{m,t}$ is replaced by $\hat{\hbf}_{m,t}$, $\forall m$, in constraints \eqref{P9:C2} and \eqref{P9:C3}.

\subsubsection{\textbf{Final Scaling of Receive Beamforming and Transmit Weights}}\label{optimization-step3-IMCSI}

Similar to Section \ref{optimization-step3}, under the solution    ($\mathbf{f}^\text{\tiny AO}_t,\{a^\text{\tiny AO}_{m,t}\})$ obtained via alternating optimization, constraint \eqref{ImperfectCSIProblem:c2} may not be satisfied with equality. In this case, we  scale the solution as in \eqref{scaling} to further reduce the sum power objective. 
The scaling factor $p_t$ in this case is given by
\begin{align}
\label{scalingfactor-IMCSI}
p_t  = \frac{1}{\| \mathbf{f}^\text{\tiny AO}_t\|}\frac{\sqrt{2}K}{\sqrt{D} \sigma_n}\sqrt{(\delta-\alpha^2) V_t^2 + \beta- \zeta },
\end{align}
where $\zeta \triangleq  \frac{\epsilon D \| \mathbf{f}^\text{\tiny AO}_t\|^2 }{ 2K^2}\sum_{m} \sigma_{h_m}^2 \left|a^\text{\tiny AO}_{m,t}\right|^2$.

\subsection{{Initialization Method}}\label{optimization-step4-IMCSI}
We adopt the initialization approach similar to that in Section~\ref{Init} to find an initial feasible point $\fbf_t$ for problem \eqref{ImperfectCSIProblem}. First, we  apply \eqref{Zeroforcing}  to set the first term on the LHS of \eqref{ImperfectCSIProblem:c2} to zero by replacing $\hbf_{m,t}$ with $\hat{\hbf}_{m,t}$. 
Next, to expand the feasible set of $\{a_{m,t}\}$, we aim to minimize the  second term $\frac{ D \| \mathbf{f}_t\|^2}{ 2K^2}\big(\sigma_n^2+ \epsilon\sum \limits_{m=1}^M \sigma_{h_m}^2 \left|a_{m,t}\right|^2\big)$ on the LHS of in \eqref{ImperfectCSIProblem:c2}, while satisfying \eqref{ImperfectCSIProblem:c3}. However, under estimated CSI,  $a_{m,t}$ appears in the above expression, and using  \eqref{Zeroforcing}, the optimization w.r.t. $\fbf_t$ will be challenging to solve.
Note that,
using \eqref{ImperfectCSIProblem:c3}, we can bound  $\sigma_n^2 + \epsilon\sum \limits_{m=1}^M \sigma_{h_m}^2 \left|a_{m,t}\right|^2 \le\sigma_n^2 + \epsilon P_0 \sum \limits_{m=1}^M \sigma_{h_m}^2$. Thus, we use the above relation to minimize the upper bound on the second term w.r.t. $\fbf_t$ to enlarge the feasible set  of $\{a_{m,t}\}$. This leads to the same optimization problem as in problem \eqref{feasiblityProb} by replacing  $\hbf_{m,t}$ with $\hat{\hbf}_{m,t}$, and can be solved in the same manner.   

The overall computational complexity of the proposed algorithm remains the same as that of the perfect CSI case, which is $\mathcal{O}((M+N)^3)$.

\section{Simulation Results}

We evaluate the efficacy of our proposed methods for image classification, where we consider the training the logistic regression for MNIST \cite{MNIST} and the convolutional neural network (CNN) for CIFAR-$10$ \cite{CIFAR10}. Note that although the loss function for CNN is non-convex, we will show that our proposed methods are still effective. 

We consider $M=10$ devices and $N=16$ antennas at the server. The distance of device $m$ from the server is randomly generated, i.e., $d_m \sim \text{Uniform}[10,100]$ meters. The path loss follows the COST Hata model \cite{COSTHATA}, i.e., $\textit{PL} 
 \text{[dB]} = 139.1 + 35.22 \log_{10}(d_m[\text{km}])$. The channel vector for device $m$ is constant during the training and is generated using a complex Gaussian distribution as $\mathbf{h}_{m, t} = \mathbf{h}_{m}  \sim \mathcal{CN}(\mathbf{0}, \frac{1}{PL}\mathbf{I})$. 
The thermal noise power is 
$N_0 =-114$ dBm, which assumes bandwidth \( \Delta f = 100 \, \text{kHz} \), and a noise figure of 10 dB. We set the total noise power \( \sigma_n^2 = -74 \, \text{dBm} \), which takes into account of both the additional interference and the thermal noise at the receiver.

We refer to  FL using our proposed joint beamforming algorithm and initialization method in Sections~\ref{PoFL} and \ref{Init} as \textit{Power Minimizing FL (PoMFL)}, and FL using the proposed beamforming design under estimated CSI in Section~\ref{impsection} as \textit{Power Minimizing FL with Imperfect CSI (PoMFL-ImCSI)}. For comparison, we consider the following three approaches as benchmarks:\\
\textbf{1) Minimum Mean Square Error (MMSE):} This approach aims to jointly optimize $\{ a_{m,t}\}$ and $\mathbf{f}_t$ to minimize the upper bound of mean squared error (MSE) given in \eqref{Proposition2:eq} under the transmit power constraint, in each communication round. To solve this, we note that $a_{m,t}$ in \eqref{Zeroforcing} minimizes the first term of the upper bound, and the minimization of the second term under the transmit power constraint reduces to problem \eqref{feasiblityProb} in our initialization method and can be solved accordingly. \\
\textbf{2) Greedy Spatial Device Selection (GSDS)\cite{MyPaper}:} GSDS is an iterative approach that jointly optimizes receive beamforming and device selection to minimize an upper bound on the optimality gap of the global loss function after $T$ rounds, where the device selection is performed iteratively in a greedy manner based on the level of channel correlation with the existing selected devices.\\
\textbf{3) Power Minimization with Bounded MSE (Bounded MSE):} 
Since MSE is used in the existing methods~  \cite{yang2020federated, liu2021reconfigurable, Paper2023}, we consider the problem of minimizing the sum device power while ensuring the MSE of the received signal is bounded by a threshold. This problem is formulated as 
\begin{subequations}\label{BoundedMSE}
    \begin{align}
    \min_{\mathbf{f}_t, \{a_{m,t}\}} \quad & \sum_{m=1}^M \left|a_{m,t}\right|^2 \label{P5:objective}\\
    \textrm{s.t.} \quad
    & \frac{D}{K^2} \Big(\sum_{m=1}^M \big|K_m v_{m,t}- \mathcal{R}[\mathbf{f}_t^H \mathbf{h}_{m, t} a_{m,t}]\big| \Big)^2 \nonumber\\ & \qquad \qquad \qquad + \frac{ D \sigma_n^2 \| \mathbf{f}_t\|^2}{ 2K^2}  \le \eta, \label{BoundedMSE:C1}\\
    & \left|a_{m,t}\right|^2 \le P_0, \: \forall m, \label{BoundedMSE:C2}
    \end{align}
\end{subequations}
where the LHS of \eqref{BoundedMSE:C1} is the upper bound of MSE given by \eqref{Proposition2:eq},  $\eta$ is the MSE threshold, and \eqref{BoundedMSE:C2} is the average transmit power constraint for each device. To solve problem \eqref{BoundedMSE}, we use the alternating optimization approach similar to the one proposed in Section \ref{PoFL}.

We point out that the average transmit power constraint \eqref{BoundedMSE:C2} is imposed  in all methods   for fair comparison. Below, we first compare the performance under the perfect CSI, and then study the performance under imperfect CSI in Section~\ref{simu:imperfectCSI}.

\subsection{Performance under MNIST Dataset}
In the MNIST dataset, each data sample is a labeled grey-scaled handwritten digit image of size $28 \times 28 $ pixels, i.e., $\mathbf{x}_k \in \mathbb{R}^{784}$, with a label $y_k \in \{0,1,...,9 \}$ to indicate its class. There are $60000$ training and $10000$ test samples. We consider training a multinomial logistic regression classifier with cross-entropy loss. The model parameters for each class consist of 784 weights and a bias term. We use a regularization term for the global loss function as $ \frac{\mu}{2} \| \mathbf{w}\|^2$ where $\mu$ is the regularization constant and set to $\mu = 10^{-4}$.
Each device's local dataset contains an equal number of
data samples from different classes, and data samples are i.i.d.  distributed among devices with uniform distribution. The number of data samples in each local dataset is $K_m =5420$. All methods employ the full local batch for gradient computation. Each (training) epoch has one training pass over the entire training dataset by all devices and thus, consists of a single communication round. 

For performance comparison, we set a target accuracy of $85\%$ and stop the training process once the test accuracy reaches the target.
For MMSE and GSDS, the learning rate is set to  $0.25$,  which is obtained after tuning to achieve the optimal performance. For PoMFL, the hyperparameters are $\alpha$, $\delta$, and $\beta$.  To reduce  the number of hyperparameters to be tuned, we set $\beta=0$. We tune $\alpha$ and $\delta$ along with the learning rate to reach the target accuracy within the same number of epochs as the GSDS and MMSE methods. This approach allows us to fairly compare the total transmit power used by different methods under a similar convergence rate, which is roughly measured by the accuracy attained over the epochs.  The optimal values of these parameters after tuning are set to $\alpha = 0.55$, $\delta= 6.0$, and the learning rate is set to $0.95$.
Similarly, for Bounded MSE, we tune  the MSE threshold  $\eta$ along with the learning rate to achieve the target accuracy within the same number of epochs as those for GSDS or MMSE. 
The optimal threshold after tuning is set to $\eta = 0.075$ and the learning rate is set to $0.75$.
All results are averaged over 20 channel realizations. 

\begin{figure}[!t]
\centering
\vspace*{-1.5em}
\includegraphics[width=0.45\textwidth, height=0.25\textwidth]{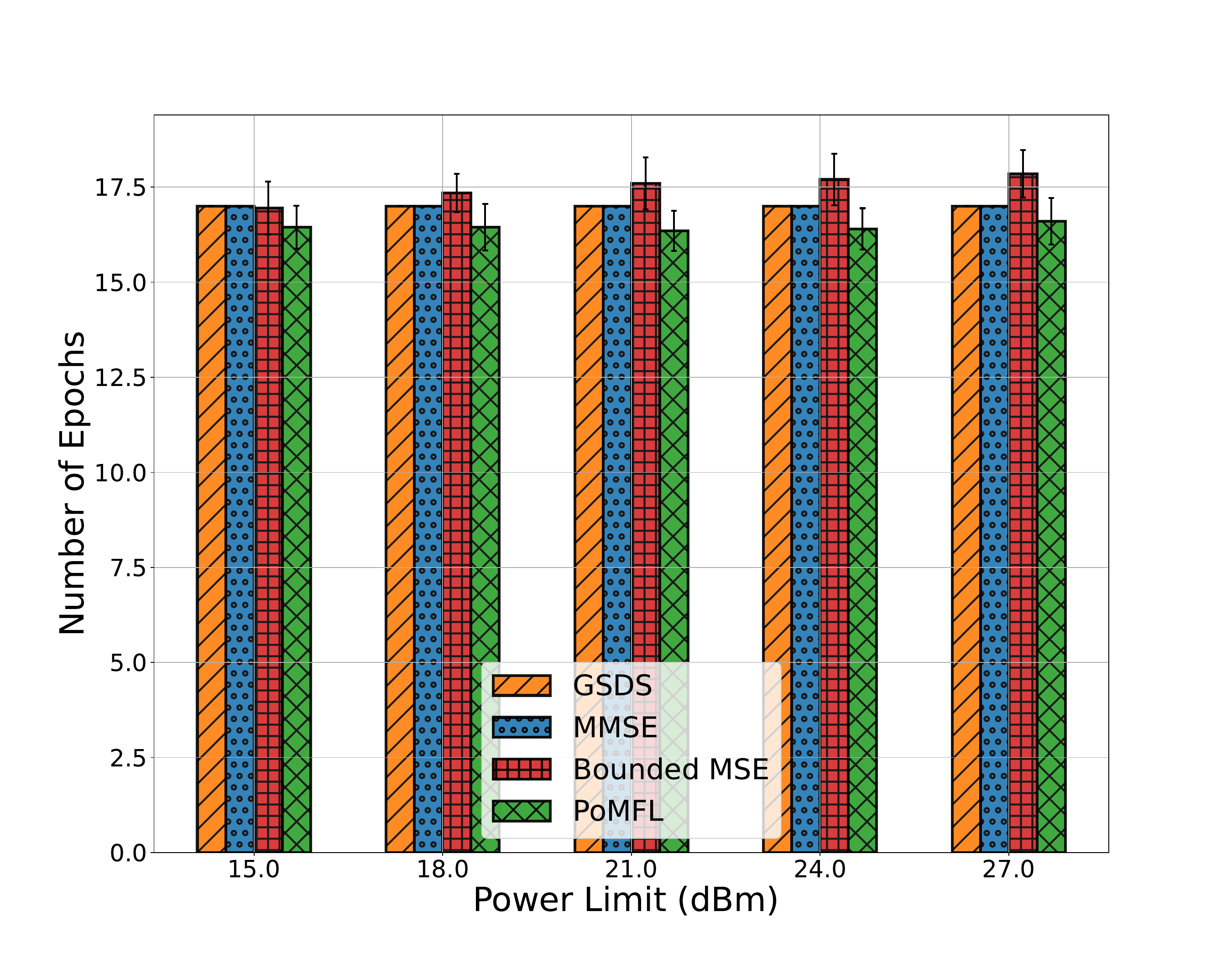}\vspace*{-1em}
\caption{Average number of epochs vs. power limit $P_0$ (MNIST).}
\label{mnist_round}
\centering
\includegraphics[width=0.45\textwidth, height=0.25\textwidth]{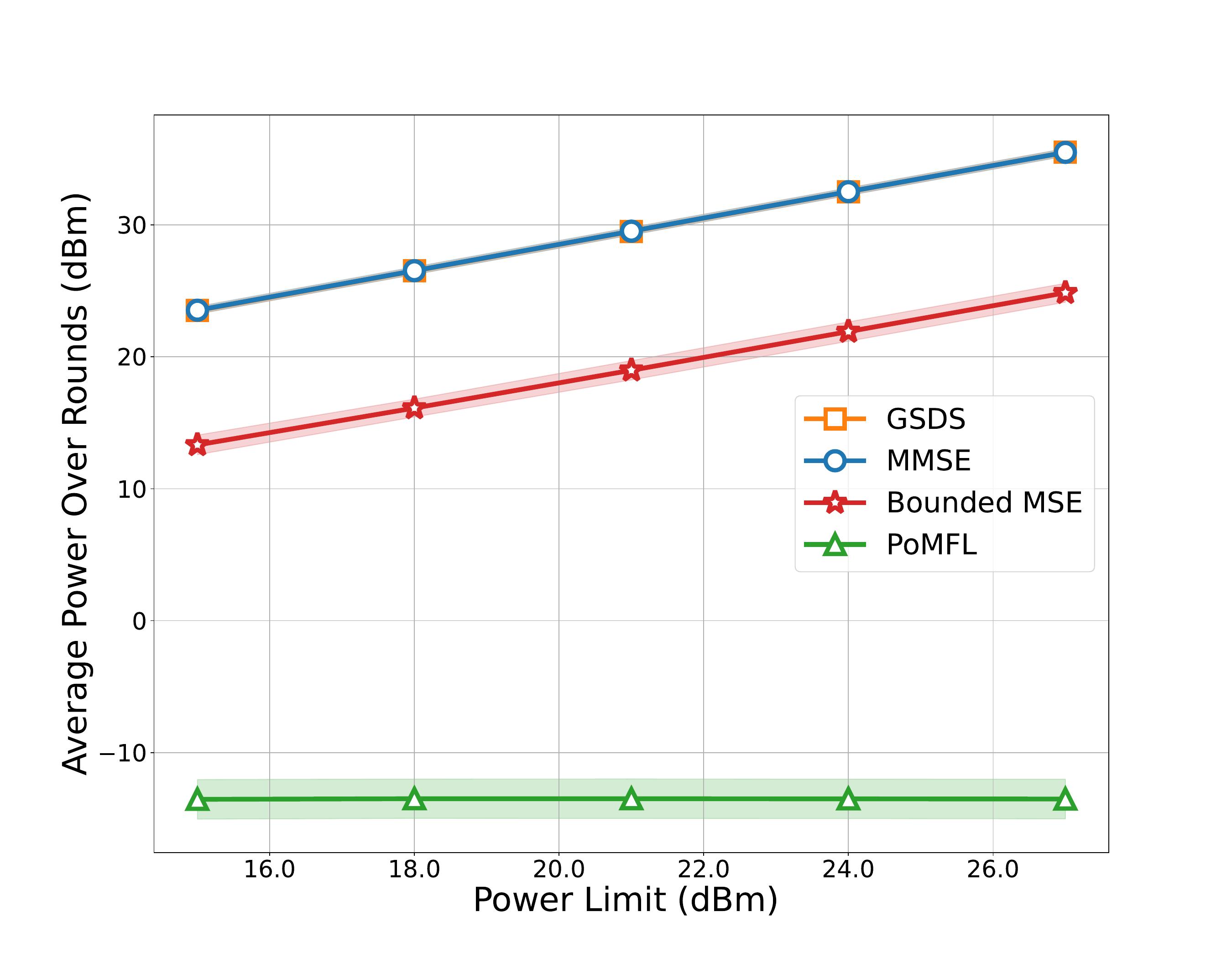}\vspace*{-1em}
\caption{Time-averaged transmit power vs. power limite $P_0$ (MNIST).}
\label{mnist_power}\vspace*{-1em}
\end{figure}

Fig.~\ref{mnist_round} shows the number of epochs needed to attain the target test accuracy by different methods at different average device transmit power limit  $P_{0}$. We consider $P_0$ in the range of 15 dBm to 27 dBm, which is the typical range for IoT devices power consumption \cite{IoTPower1, IoTPower2}.  
The error bar indicates the 95$\%$ confidence intervals. We see that most of the methods achieve the target accuracy in approximately 17 epochs, verifying that our parameter tuning for PoMFL and Bounded MSE leads to achieving the target accuracy within the same epoch as GSDS and MMSE.

Fig.~\ref{mnist_power} shows the time-averaged sum transmit power vs. power limit $P_0$. The time-averaged transmit power is given by $\frac{1}{T}\sum_{t=0}^{T-1}\sum_{m=1}^M \left|a_{m,t}\right|^2$, where $T$ is the number of communication rounds to reach the target test accuracy.
The shaded area around each curve shows the $95\%$ confidence intervals. Our proposed PoMFL significantly outperforms the benchmarks. With the same number of epochs to attain the target test accuracy as given in Fig.~\ref{mnist_round}, PoMFL reduces the total power consumption by more than $25$ dB compared to  Bounded MSE  and more than $35$ dB compared to GSDS and MMSE.
This demonstrates PoMFL\ is significantly more power efficient than the benchmarks for FL training without compromising the convergence rate.

Moreover, we notice that the time-averaged sum transmit power by PoMFL remains constant for different values of $P_0$. The reason is that PoMFL can effectively reduce the transmit power at each devices, and none of the devices operate at their maximum power limit $P_0$, i.e., the power constraint \eqref{P5:C3} is inactive, and thus increasing the power limit does not affect the sum transmit power. However, the other methods always yield the transmit power that reaches the maximum power limit $P_0$, and therefore, increasing $P_0$ leads to increased device power consumption.

\subsection{Performance under CIFAR-10 Dataset}\label{cifarSim}

In the CIFAR-10 dataset, each data sample consists of a colored image of size $3\times 32 \times 32 $ pixels, i.e., $\mathbf{x}_k \in \mathbb{R}^{3}\times \mathbb{R}^{32} \times \mathbb{R}^{32}$ and label $y_k \in \{0,1,...,9 \}$ indicates the class of the image. There are $50000$ training and $10000$ test samples. A CNN model named the Residual Network with 14 layers (ResNet-14) \cite{ResNet} is trained using the cross-entropy loss. 
Similarly to the MNIST dataset, we evenly distribute the training samples from each class among the devices with $K_m = 5000$. 
During each communication round, the devices compute their local gradient using a batch of data of size 10 from their respective local dataset, and hence each epoch consists of 500 communication rounds. We set the  SGD momentum and the weight decay to $0.9$ and $10^{-4}$, respectively for all methods. 

We set the target accuracy of $70\%$ to terminate the training process.
The learning rates for the MMSE and GSDS methods are set to $0.01$ after being tuned to achieve the optimal performance. Similar to the MNIST experiments, to ensure a fair comparison, for PoMFL, we set $\beta=0$ and tune $\alpha$ and $\delta$ along with the learning rate to reach the target accuracy within the same number of epochs as  GSDS and MMSE. After hyperparameter tuning, we set $\alpha = 0.05$, $\delta= 0.05$,  and the learning rate to $0.2$.
Similarly, for Bounded MSE, we use the same tuning objective and set $\eta = 0.4$, and the learning rate to $0.2$.
All results represent averages over 20 channel realizations. We again use either error bar or shaded area around each curve to indicate the 95$\%$ confidence intervals.  

\begin{figure}[!t]
\centering
\vspace*{-1.5em}
\includegraphics[width=0.45\textwidth, height=0.25\textwidth]{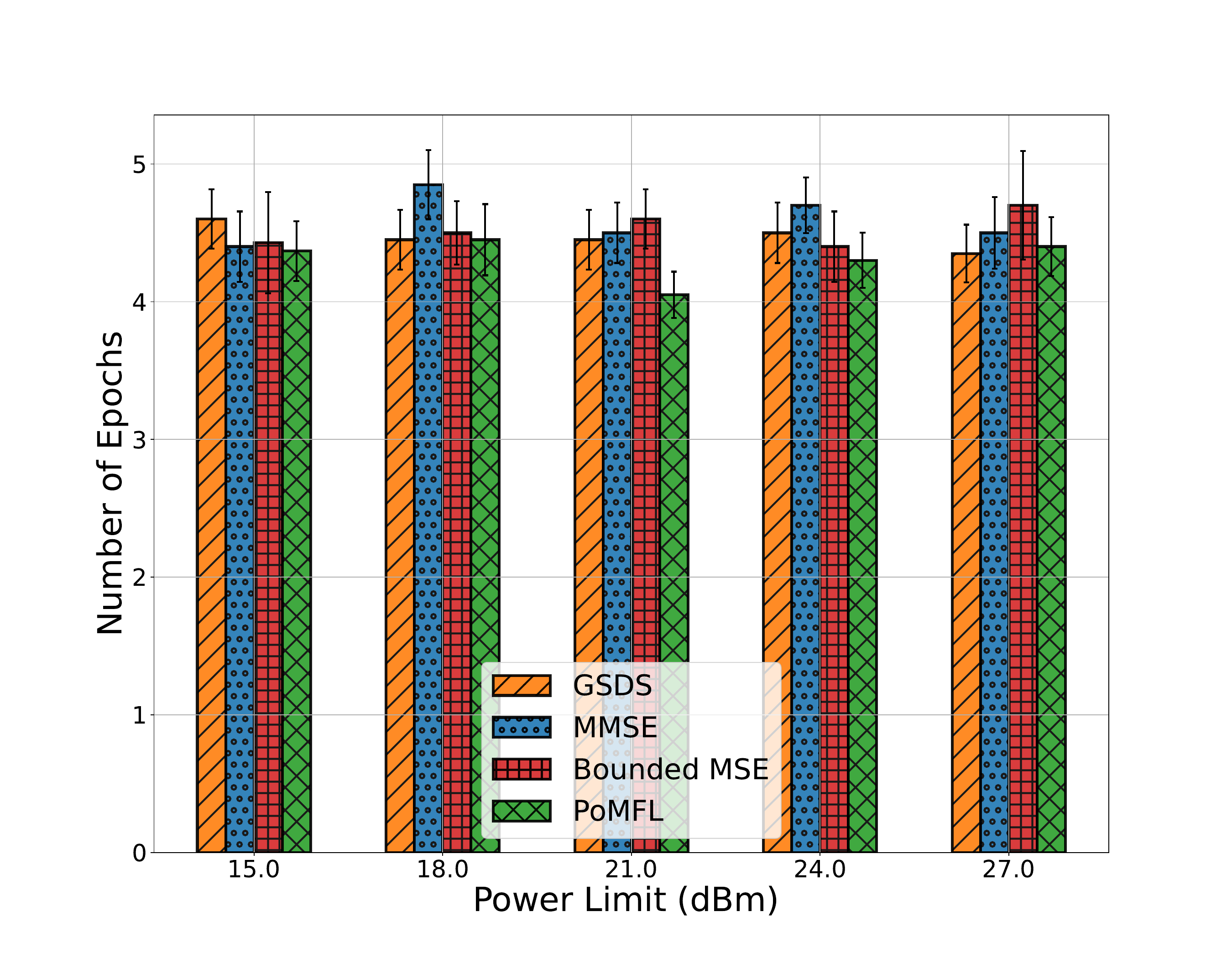}\vspace*{-1em}
\caption{Average number of epochs vs. power limit $P_0$ (CIFAR-10).}
\label{cifar_round}
\centering
\includegraphics[width=0.45\textwidth, height=0.25\textwidth]{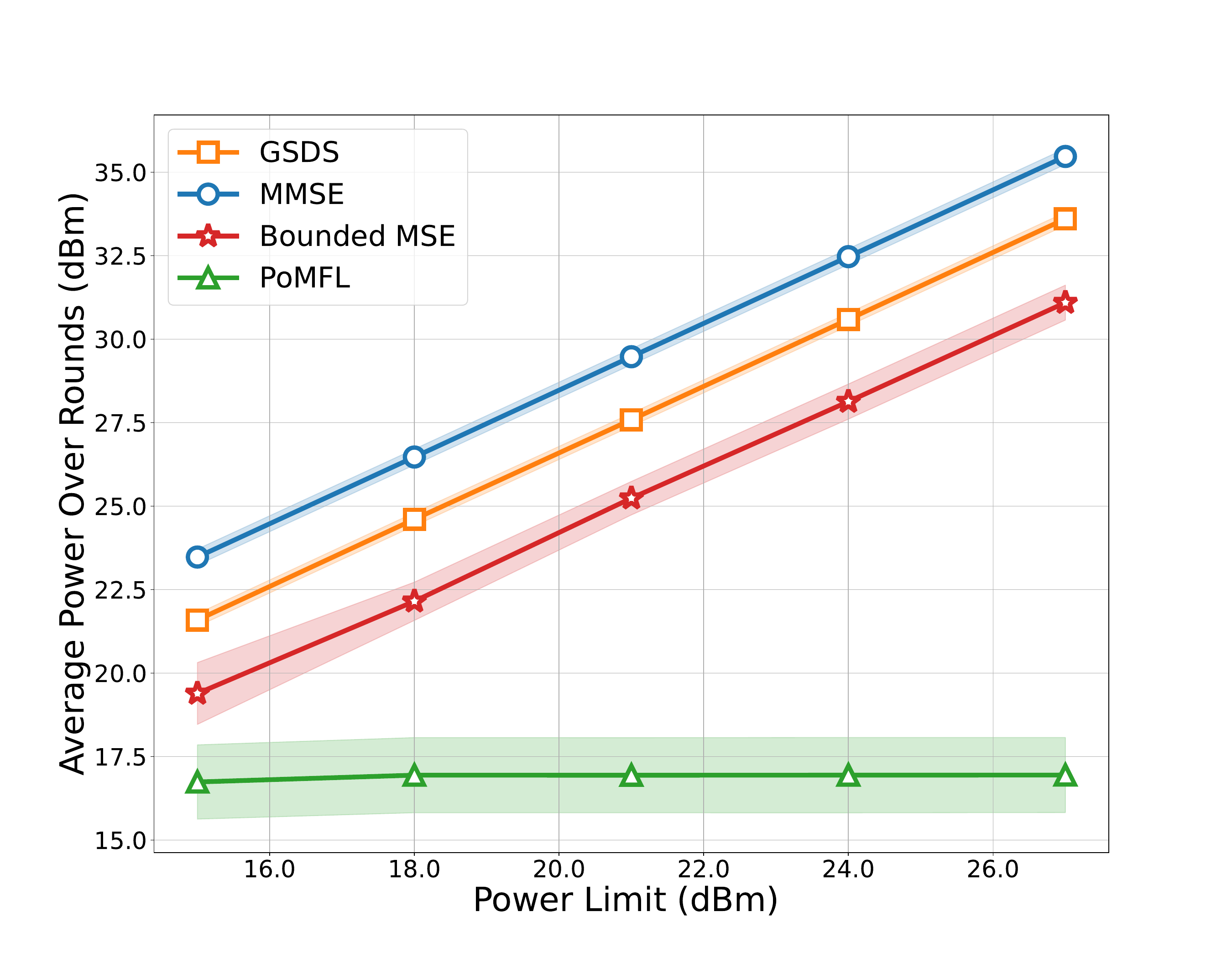}\vspace*{-1em}
\caption{Time-averaged transmit power vs. power limit $P_0$  (CIFAR-10).}
\label{cifar_power}\vspace*{-1em}
\end{figure}

Fig.~\ref{cifar_round} shows the number of epochs needed to attain the target test accuracy by different methods at different average device transmit power limits $P_{0}$. All methods achieve the target accuracy in approximately 4.5 epochs, which is intended by the parameter tuning for PoMFL and Bounded MSE.

Fig.~\ref{cifar_power} shows the time-averaged sum transmit power vs. power limit $P_0$. Similar to Fig.~\ref{mnist_power}, our proposed PoMFL again shows significant power saving over the benchmarks, with the same number of epochs to attain the target test accuracy as given in Fig.~\ref{cifar_round}.
In particular, compared to Bounded MSE, which is the best among the three benchmarks, PoMFL reduces the average total power consumption by 3 dB up to 16 dB for $P_0$ ranging from 15 dBm to 27 dBm, respectively. This demonstrates that with a more complicated training dataset,  PoMFL remains to be significantly more power efficient than the benchmarks without compromising the convergence rate.
Also,  the time-averaged sum transmit power by PoMFL remains constant as $P_0$ increases, unlike the benchmarks where the transmit power increases with $P_0$.

Comparing Figs.~\ref{mnist_power} and~\ref{cifar_power}, we see that the power saving of PoMFL is more significant for the MNIST dataset compared to CIFAR-10. 
This is because the MNIST dataset is generally easier to learn than CIFAR-10, and the model can afford a bigger error margin. This allows PoMFL to use even lower transmit power to achieve the same convergence speed.

\subsection{Performance under CIFAR-10 Dataset with Imperfect CSI}\label{simu:imperfectCSI}
We now consider the case when the server only has the imperfect CSI and compare the performance of different methods using the CIFAR-$10$ dataset. We set a target accuracy of 70$\%$. Other settings are the same as those used in Section \ref{cifarSim}. We set the power limit to 21 dBm. For PoMFL, we apply the algorithm in Section~\ref{PoFL} by simply replacing $\hbf_{m,t}$ with $\tilde{\hbf}_{m,t}$, $\forall m$. For both PoMFL and PoMFL-ImCSI, we set $\alpha=0.05$, the same as in Section \ref{cifarSim}. Additionally, we tune the parameters  of our methods and Bounded MSE  to ensure they converge to the target accuracy within the same number of epochs as GSDS and MMSE. The optimized values of these parameters for each method are listed in Table \ref{table2}.
All results represent averages over 20 channel realizations. 

\begin{figure}[!t]
\centering
\vspace*{-1em}
\includegraphics[width=0.45\textwidth, height= 0.25\textwidth]{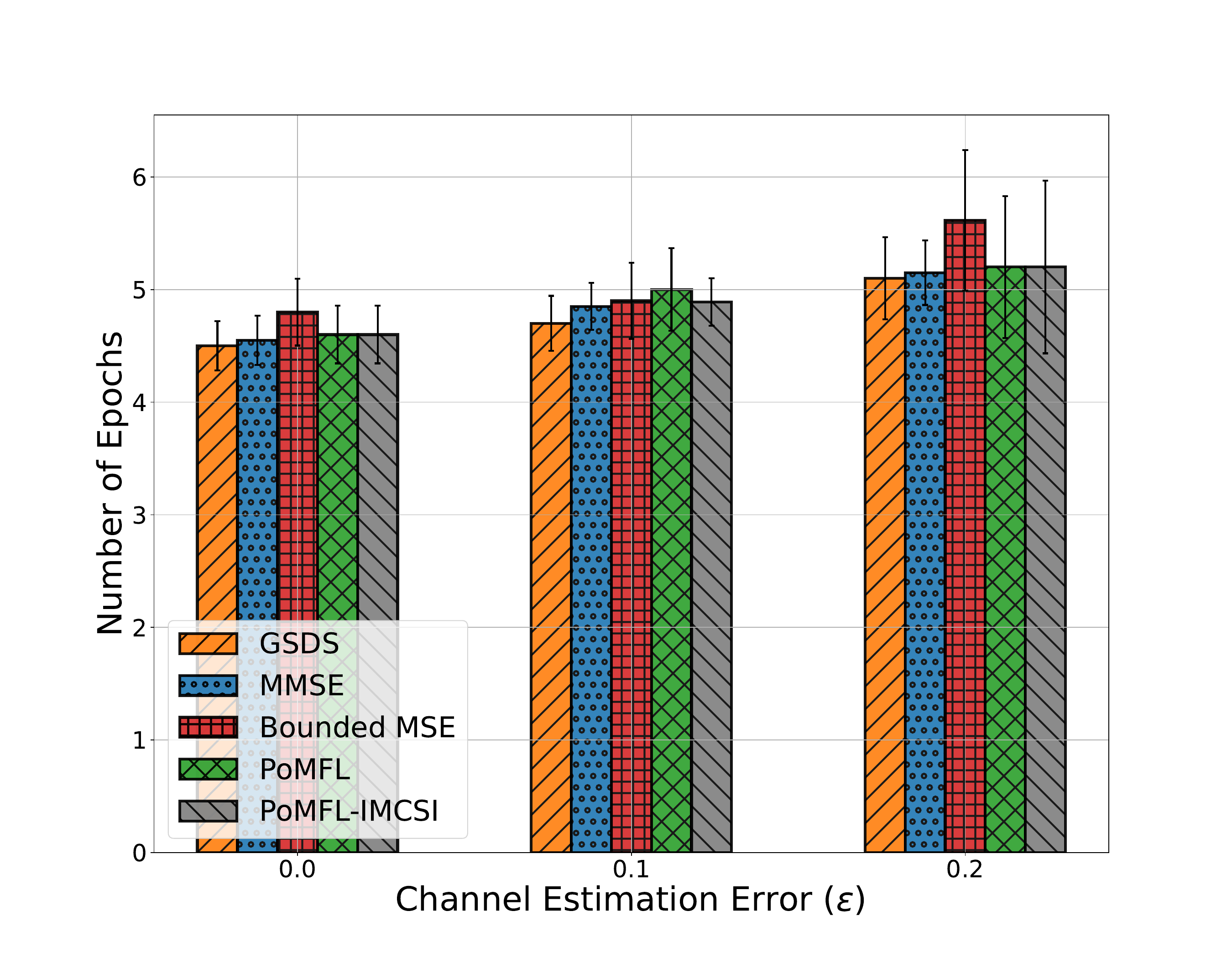}\vspace*{-1em}
\caption{Average number of epochs (CIFAR-10 with imperfect CSI).}
\label{cifar_round_imperfect}
\centering
\includegraphics[width=0.45\textwidth, height= 0.25\textwidth]{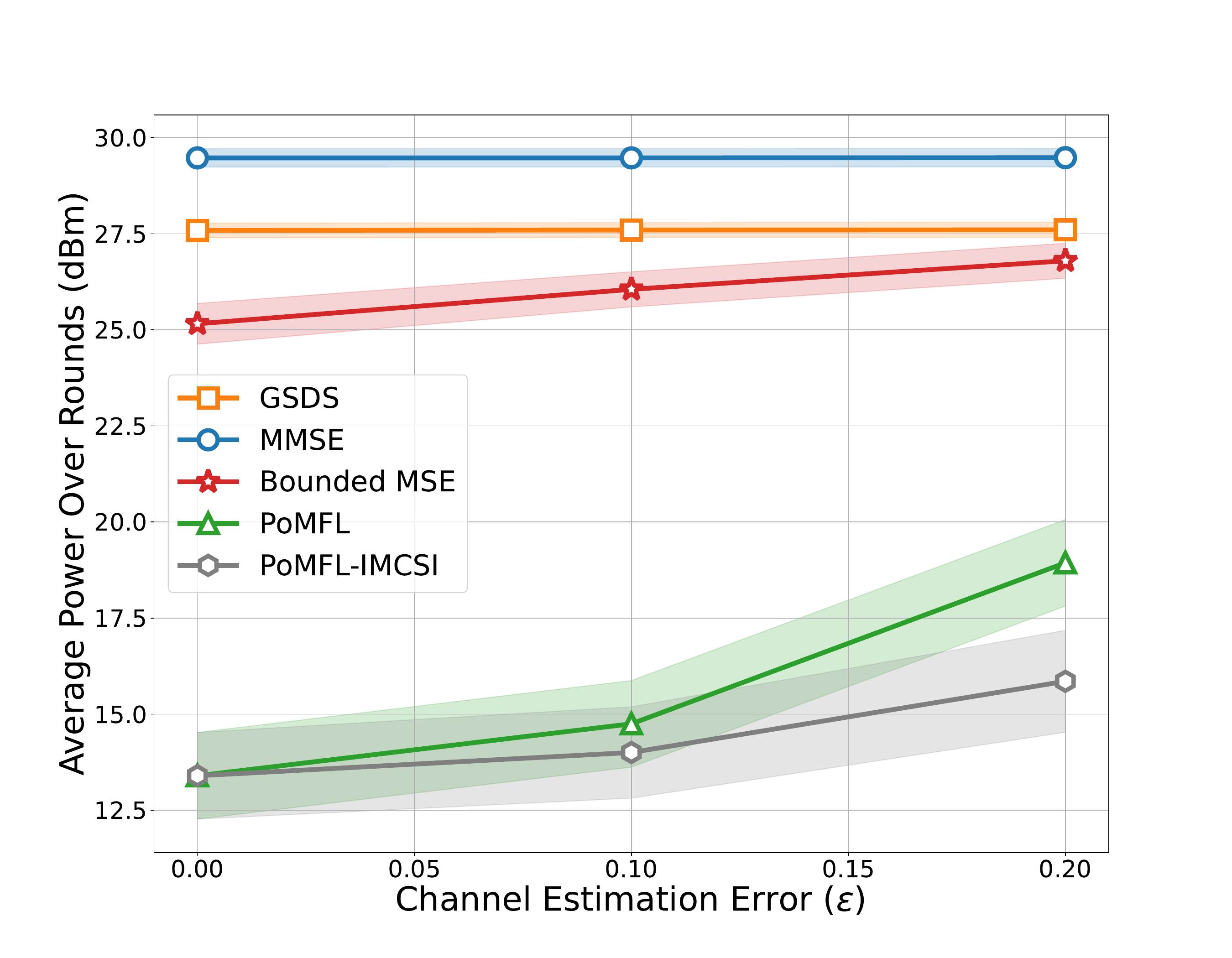}\vspace*{-1em}
\caption{Time-averaged transmit power vs. CSI error $\epsilon$ (CIFAR-10 with imperfect CSI).}
\label{cifar_power_imperfect}\vspace*{-1em}
\end{figure}

Fig.~\ref{cifar_round_imperfect} shows the average number of epochs required by each method to achieve the target accuracy for various levels of CSI error $\epsilon$. They are approximately the same as we intended for parameter tuning.
However, the number of epochs required for convergence increases as the CSI error $\epsilon$ increases, which is expected as the error causes training convergence to slow down. 

Fig.~\ref{cifar_power_imperfect} depicts the time-averaged total power by different methods under different levels of CSI error $\epsilon$. Both PoMFL and PoMFL-ImCSI still use significantly lower power than the benchmark methods at different levels of CSI error.
Compared with PoMFL, PoMFL-ImCSI takes into account of CSI error in the joint beamforming design, resulting in a lower total power than that of PoML. This gap becomes more significant as $\epsilon$ increases from 0.1 to 0.2, with power saving increases  from 0.75 dB to 3 dB.

\begin{table}[t]
\caption{\textnormal{Hyperparameters  used for different CSI error $\epsilon$}.}
\begin{center}
\begin{tabular}{|r||p{40pt}| p{40pt} |p{40pt}| }
\hline
\textbf{Method} & 
$\text{\boldmath{$\epsilon=0.0$ }}$ & $\text{\boldmath{$\epsilon=0.1$ }}$ & $\text{\boldmath{$\epsilon=0.2$ }}$ \\
\hline\hline
Bounded MSE & $\eta = 0.4$ &  $\eta = 0.25$ & $\eta = 0.15$ \\ \hline
PoMFL & $ \delta = 0.11$& $\delta = 0.09$ & $ \delta = 0.04$ \\ \hline
PoMFL-ImCSI & $ \delta = 0.11$ & $\delta = 0.11$ & $\delta = 0.09$ \\ \hline
\end{tabular} 
\label{table2}
\end{center}
\vspace*{-1em}
\end{table}

\section{Conclusion}
In this paper, we propose a power-efficient joint transmit and receive beamforming design to improve the uplink over-the-air analog aggregation for FL training. Our objective is to minimize the average total transmit power consumed among devices  while ensuring FL training convergence to the optimal model. We provide sufficient conditions on the expected aggregation error per communication round to guarantee FL convergence and formulate a transmit power minimization problem to jointly optimize the transmit weights and receive beamforming.
Our proposed PoMFL algorithm solves this problem using alternating optimization which guarantees convergence to a partial optimum. We also propose an effective initialization method that significantly accelerates the convergence. We further propose an CSI-error-aware joint beamforming algorithm PoMFL-ImCSI when only estimated CSI is available, which utilizes the estimated CSI and the CSI error to determine the transmit weights and receive beamforming. Simulation on different training datasets and learning models shows that our proposed algorithms are highly power-efficient and use significantly lower power than the existing methods to attain the same target accuracy under the same convergence rate.

\appendices 
\section{Proof of Theorem 1}
\label{Appendix:A}
\IEEEproof
We use the following theorem to prove our results.
\begin{theorem}
\cite[Chapter 2, Theorem 3]{PolyakBook}
\label{polyaktheorem}
    Assume the following conditions hold:
    \begin{enumerate}
        \item The distribution of $\mathbf{s}_t$ depends only on $\mathbf{w}_t$ and $t$, and given the sequence $\{\mathbf{w}_t\}$, the sequence $\{\mathbf{s}_t\}$ is independent.
        \item There is a scalar Lyapunov function $V(\cdot) \ge 0$ that is differentiable and $\nabla V(\cdot)$ is $L$-Lipschitz continuous.
        \item Process $\mathbf{s}_t$ is pseudogradient in relation to $V(\mathbf{w}_t)$, i.e., \\ $ \big \langle \nabla V(\mathbf{w}_t), \mathbb{E}[(\mathbf{s}_t|\mathbf{w}_t)] \big \rangle \ge b V(\mathbf{w}_t)$, where $b >0 $ is a constant scalar and $\langle \cdot, \cdot \rangle $ represents the inner product.
        \item The following growth condition on $\mathbf{s}_t$ is satisfied: $ \mathbb{E}[\|\mathbf{s}_t\|^2| \mathbf{w}_t] \le \sigma^2+ \tau \big \langle \nabla V(\mathbf{w}_t), \mathbb{E}[(\mathbf{s}_t|\mathbf{w}_t)] \big \rangle$,
        where $\tau \ge 0$ and $\sigma^2$ are two constants.
        \item The initial point satisfies $\mathbb{E}[V(\mathbf{w}_0)] < \infty$.
        \item The learning rate is such that: $
            \gamma_t \ge 0, \sum_{t=0}^\infty \gamma_t = \infty, \underset{t \to \infty}{\limsup} \: \gamma_t < \frac{2}{L \tau}, \forall t.$
    \end{enumerate}
    Let either $\sigma^2=0$ or $\underset{t \to \infty}{\lim} \gamma_t = 0$. Then the gradient descent method in \eqref{eq12} results in $ \underset{t \to \infty}{\lim} \mathbb{E} [V(\mathbf{w}_t)] = 0$.
\end{theorem}

Consider the Lyapunov function $V(\mathbf{w}) = F(\mathbf{w}) - F(\mathbf{w}^*)$. We will show that under assumptions \textbf{A1}-\textbf{A3} and conditions \textbf{C1}-\textbf{C3}, the conditions stated in Theorem 2 are satisfied, which directly implies our theorem statement. From the definition of $\mathbf{s}_t$, we have
\vspace{-0.5em}
\begin{align}\label{Theorem3-1}
\mathbf{s}_t & = \frac{1}{K} \sum_{m=1}^M\Big( \frac{\mathcal{R}[\mathbf{f}^{H}_t  \mathbf{h}_{m, t} a_{m,t}]}{v_{m,t}}\Big)\mathbf{g}_{m,t} + \frac{\tilde{\mathbf{n}}_t}{K}.
\end{align} 

Given $\mathbf{w}_t$, $\mathbf{s}_t$ is only a function of $\{\mathbf{n}_{d,t}\}$. Since $\{\mathbf{n}_{d,t}\}$ is independent over $t$ and also independent of $\mathbf{w}_t$, the values of $\mathbf{s}_t$ are independent of each other over $t$. Hence, given $\mathbf{w}_t$, the first condition in Theorem~\ref{polyaktheorem} is satisfied.

Since $V(\mathbf{w})$ is a difference of two $L$-Lipschitz continuous functions, it is also $L$-Lipschitz continuous, and thus the second condition in Theorem~\ref{polyaktheorem} is satisfied.

For the third condition of Theorem~\ref{polyaktheorem}, we have
\begin{align}\label{Theorem3-2}
\big \langle \nabla V(\mathbf{w}_{t}), \mathbb{E}[\mathbf{s}_{t} | \mathbf{w}_{t}] \big \rangle &  \overset{\mathrm{(a)}}{=}\nonumber \big\langle \:\nabla F(\mathbf{w}_{t}), \mathbb{E}[\nabla F(\mathbf{w}_t)+ \mathbf{e}_{t} | \mathbf{w}_{t}] \: \big\rangle \nonumber \\ &  \hspace{-5em} \overset{\mathrm{(b)}}{\ge} \| \nabla F(\mathbf{w}_{t}) \|^2-\!\| \nabla F(\mathbf{w}_{t}) \|  \|\mathbb{E}[\mathbf{e}_{t} | \mathbf{w}_{t}]\| \nonumber\\ & \hspace{-5em} \overset{\mathrm{(c)}}{\ge} (1-\alpha) \| \nabla F(\mathbf{w}_{t}) \|^2  \overset{\mathrm{(d)}}{\ge} 2\mu (1-\alpha) V(\mathbf{w}_t),
\end{align}
where (a) follows the definition of $\mathbf{s}_t$ in \eqref{UpdateDef} and the fact that $\nabla V(\mathbf{w})= \nabla F(\mathbf{w})$. Inequality 
(b) follows the Cauchy–Schwarz Inequality, and (c) follows our assumption in \textbf{C1}. Inequality (d) follows the fact that for strongly convex functions $\| \nabla F(\mathbf{w}_{t}) \|^2 \ge 2 \mu (F(\mathbf{w}_t)-F(\mathbf{w}^*))= 2 \mu V(\mathbf{w}_t)$. So, the third condition of Theorem~\ref{polyaktheorem} holds with $b= 2\mu (1-\alpha)$. 

For the fourth condition of Theorem~\ref{polyaktheorem}, based on the definition of $\mathbf{s}_t$ in \eqref{UpdateDef}, we have 
\begin{align}\label{Theorem3-3}
 \mathbb{E}[& \| \mathbf{s}_{t} \|^{2} |\mathbf{w}_{t}] \nonumber \\ & \overset{\mathrm{(a)}}{=} 
\| \nabla F(\mathbf{w}_{t})\|^2 + 2 \nabla F(\mathbf{w}_{t})^T \mathbb{E}[\mathbf{e}_{t}| \mathbf{w}_t] + \mathbb{E}[\| \mathbf{e}_{t}\|^2| \mathbf{w}_t] \nonumber \\ &   \overset{\mathrm{(b)}}{\le}  \| \nabla F(\mathbf{w}_{t})\|^2 \!+ 2 \nabla F(\mathbf{w}_{t})^T \mathbb{E}[\mathbf{e}_{t}| \mathbf{w}_t] +  \delta \| \nabla F(\mathbf{w}_{t})\|^2+ \! \beta \nonumber\\ &  \overset{\mathrm{(c)}}{\le}  (1+\delta) \| \nabla F(\mathbf{w}_{t})\|^2 +  2 \| \nabla F(\mathbf{w}_{t})\| \|  \mathbb{E}[\mathbf{e}_{t}| \mathbf{w}_t]\|+\beta \nonumber\\ &  \overset{\mathrm{(d)}}{\le}  (1+\delta+2\alpha) \| \nabla F(\mathbf{w}_{t})\|^2 +\beta \nonumber \\ & \overset{\mathrm{(e)}}{\le} (1+\delta+2\alpha)  \frac{\big \langle \nabla V(\mathbf{w}_{t}), \mathbb{E}[\mathbf{s}_{t} | \mathbf{w}_{t}] \big \rangle}{(1-\alpha)} +\beta,
\end{align}
where (b) follows our assumption in \textbf{C2}, (c) follows the Cauchy-Schwarz Inequality, (d) follows our assumption in \textbf{C1}, and finally (e) follows inequality (c) in \eqref{Theorem3-2}. Therefore, the fourth condition is satisfied with $\sigma^2= \beta$ and $ \tau = \frac{1+\delta +2 \alpha}{1-\alpha}$.

The last two conditions of Theorem~\ref{polyaktheorem} are satisfied since the loss function at the starting point is bounded, and \textbf{C3} guarantees the bounds on the learning rate outlined in the theorem.
\endIEEEproof

\section{Proof of Proposition 1}\label{Appendix:B}
\IEEEproof
According to \eqref{ErrorDef}, $\mathbb{E}[\mathbf{e}_t| \mathbf{w}_t]$ equals the first term at the RHS of \eqref{ErrorDef}
since $\mathbf{n}_{d,t}$ is zero-mean. Therefore,
\begin{align}\label{Proposition1Proof-2}
    \| \mathbb{E}[\mathbf{e}_t| \mathbf{w}_t] \| & \overset{\mathrm{(a)}}{\le} \frac{1}{K} \sum_{m=1}^M \big|\frac{\mathcal{R}[\mathbf{f}^{H}_t  \mathbf{h}_{m, t} a_{m,t}]}{v_{m,t}}- K_m \big| \|\mathbf{g}_{m,t}\| \nonumber \\
    & \overset{\mathrm{(b)}}{=} \frac{\sqrt{D}}{K} \sum_{m=1}^M \big| \mathcal{R}[\mathbf{f}^{H}_t  \mathbf{h}_{m, t} a_{m,t}]- K_m v_{m,t}\big|,
\end{align}
where (a) follows the Triangle Inequality and (b) follows the definition of $v_{m,t}$.
\endIEEEproof

\section{Proof of Proposition 2}
  \label{Appendix:C}
  \IEEEproof
Since $\mathbf{n}_{d,t} \sim \mathcal{CN}(\mathbf{0}, \sigma_n^2 \mathbf{I})$, we have
\begin{align}
&\mathbb{E}[\mathcal{R}[\mathbf{n}_{d,t}]\mathcal{R}[\mathbf{n}_{d,t}]^T]  = \mathbb{E}[\mathcal{I}[\mathbf{n}_{d,t}] \mathcal{I}[\mathbf{n}_{d,t}]^T]= \frac{\sigma_n^2}{2}\mathbf{I}, \\ & \mathbb{E}[ \mathcal{R}[\mathbf{n}_{d,t}] \mathcal{I}[\mathbf{n}_{d,t}]^T]= \mathbb{E}[ \mathcal{I}[\mathbf{n}_{d,t}]\mathcal{R}[\mathbf{n}_{d,t}]^T] = \mathbf{0}.
\end{align}
Let $\tilde{{n}}_t[d]$ denotes $d$-th entry of $\tilde{\mathbf{n}}_t$, its variance is given by
$\mathbb{E}[ \tilde{{n}}_t[d]^2]= \mathbb{E}[ \mathcal{R}[\mathbf{f}^{H}_{t} \mathbf{n}_{d,t}]^2] = \frac{\sigma_n^2}{2} \|\mathbf{f}_t\|^2,
$
and subsequently, 
$\mathbb{E}[\|\tilde{\mathbf{n}}_t\|^2] = \frac{D \sigma_n^2 \| \mathbf{f}_t \|^2}{2}$. Define $ \mathbf{u}_t \triangleq \frac{1}{K} \sum_{m=1}^M \Big( \frac{\mathcal{R}[\mathbf{f}^{H}_t  \mathbf{h}_{m, t} a_{m,t}]}{v_{m,t}}- K_m\Big)\mathbf{g}_{m,t} $.
Based on the definition of $\mathbf{e}_t$ in \eqref{ErrorDef},
\begin{align}\label{Proposition2Proof-3}
& \mathbb{E}[\|\mathbf{e}_t\|^2| \mathbf{w}_t] \nonumber \\ & \overset{\mathrm{(a)}}{=}  \| \mathbf{u}_t \|^2 +  \mathbf{u}_t^H \mathbb{E}\Big[ \frac{{\tilde{\mathbf{n}}}_t}{K} \Big]+ \mathbb{E}\Big[ \frac{{\tilde{\mathbf{n}}}^H_t}{K}\Big] \mathbf{u}_t+\mathbb{E}\Big[\frac{\|\tilde{\mathbf{n}}_t\|^2}{K^2}\Big] \nonumber\\ & \overset{\mathrm{(b)}}{=} \| \mathbf{u}_t \|^2 + \frac{D \sigma_n^2 \| \mathbf{f}_t \|^2}{2 K^2} \nonumber \\& \overset{\mathrm{(c)}}{\le} \Big( \frac{1}{K} \sum_{m=1}^M \Big| \frac{\mathcal{R}[\mathbf{f}^{H}_t  \mathbf{h}_{m, t} a_{m,t}]}{v_{m,t}}- K_m\Big|\: \|\mathbf{g}_{m,t}\|\Big)^2 + \frac{D \sigma_n^2 \| \mathbf{f}_t \|^2}{2 K^2} \nonumber\\ & \overset{\mathrm{(d)}}{=}
\frac{D}{K^2} \Big( \sum_{m=1}^M \Big| \mathcal{R}[\mathbf{f}^{H}_t  \mathbf{h}_{m, t} a_{m,t}]- K_m v_{m,t}\Big| \: \Big)^2 + \frac{D \sigma_n^2 \| \mathbf{f}_t \|^2}{2 K^2},
\end{align}
where (a) follows from expanding the expression of $\|\mathbf{e}_t\|^2$ into four terms and the fact that given $\mathbf{w}_t$, $\mathbf{u}_t$ is deterministic, and $\tilde{\mathbf{n}}_t$ is independent of $\mathbf{w}_t$; (b) follows the fact that $\tilde{\mathbf{n}}_t$ is zero-mean and substituting its variance into the last term; (c) follows applying the Triangle Inequality to $\|\mathbf{u}_t\|$; and finally (d) follows the definition of $v_{m,t}$.
\endIEEEproof

\section{Proof of Proposition 3}
\label{Appendix:D}  
\IEEEproof
The proof is by contradiction. Let's denote $\phi_{m,t} \triangleq \angle a_{m,t}^*$ and $ \psi_{m,t} =  \angle \mathbf{f}_t^H \mathbf{h}_{m, t}$. Suppose $\phi_{m,t}\neq - \psi_{m,t}$.
Let's define another set of transmit weights $\tilde{a}_{m,t} \triangleq |a_{m,t}^*| \text{cos}(\phi_{m,t}+ \psi_{m,t})  e^{-j\psi_{m,t}}, \forall m$. We have
\begin{align}\label{contradict-2}
    \mathcal{R}[{\tilde{a}_{m,t} \mathbf{f}_t^H \mathbf{h}_{m, t}}] & = |a_{m,t}^*| \text{cos}(\phi_{m,t}+ \psi_{m,t}) |\mathbf{f}_t^H \mathbf{h}_{m, t}| \\ & = \mathcal{R}[{a_{m,t}^* \mathbf{f}_t^H \mathbf{h}_{m, t}}], \forall m.
\end{align}

Therefore, $\{\tilde{a}_{m,t}\}$ can satisfy constraint \eqref{P7:C1}. Moreover, $|\tilde{a}_{m,t}| = |a_{m,t}^*| \text{cos}(\phi_{m,t}+ \psi_{m,t}) < |a_{m,t}^*|, \forall m$ and hence $\{\tilde{a}_{m,t}\}$ satisfies \eqref{P7:C2} with a lower value for the objective function. This contradicts the optimality of $\{a_{m,t}^*\}$.
\endIEEEproof

\section{Proof of Proposition 4}\label{Appendix:E}
\IEEEproof
Let $\{b_{m,t}^\star\}$ be the optimal solution of \eqref{P10}. We first show that $\{b_{m,t}^\star\}$ satisfy:
\begin{align}\label{Proposition4-1}
    b_{m,t}^\star \le \frac{K_m v_{m,t}}{|\mathbf{f}_t^H \mathbf{h}_{m,t}|}, \forall m.
\end{align}
We prove it by contradiction. Assume \eqref{Proposition4-1} does not hold for the optimal solution of \eqref{P10}, i.e., $\exists j, \frac{K_j v_{j,t}}{|\mathbf{f}_t^H \mathbf{h}_{j,t}|} \le b_{j,t}^\star. $ We define $\tilde{b}_{j,t} \triangleq \text{max}(0, \frac{2K_j v_{j,t}}{|\mathbf{f}_t^H \mathbf{h}_{j,t}|}- b_{j,t}^\star).$
We show that the set of variables 
$\{ \tilde{b}_{j,t}, b_{i,t}^\star: i \neq j \}$ satisfies all constraints of problem \eqref{P10} and it results in lower objective value and hence a contradiction of the assumption $\{ b_{i,t}^\star\}$ is the optimal solution.

We consider two cases: i) $ b_{j,t}^\star \le \frac{2K_j v_{j,t}}{|\mathbf{f}_t^H \mathbf{h}_{j,t}|}$, for which we have
\begin{align}
        \big| K_j v_{j,t} - |\mathbf{f}_t^H \mathbf{h}_{j,t}| &\tilde{b}_{j,t} \big|  = \big||\mathbf{f}_t^H \mathbf{h}_{j,t}| b_{j,t}^\star -K_j v_{j,t} \big|,
\end{align}
which is obtained by substituting $\tilde{b}_{j,t} = \frac{2K_j v_{j,t}}{|\mathbf{f}_t^H\mathbf{h}_{j,t}|}- b^\star_{j,t}.$ This indicates that the $j$-th term within the summation on the LHS of \eqref{P10:C1} remains unchanged and hence \eqref{P10:C1} holds. 
ii) $ \frac{2K_j v_{j,t}}{|\mathbf{f}_t^H \mathbf{h}_{j,t}|} \le b_{j,t}^\star$, and hence $\tilde{b}_{j,t}=0$. We have
\begin{align}
\hspace*{-0.9em}  
        \big| K_j v_{j,t} - |\mathbf{f}_t^H \mathbf{h}_{j,t}| \tilde{b}_{j,t} \big|  & =\! K_j v_{j,t} \le \! \big||\mathbf{f}_t^H \mathbf{h}_{j,t}| b_{j,t}^\star -K_j v_{j,t} \big|,
\end{align}
where the inequality holds because $ \frac{2K_j v_{j,t}}{|\mathbf{f}_t^H \mathbf{h}_{j,t}|} \le b_{j,t}^\star$. This indicates that the 
$j$-th term in the summation on the LHS of \eqref{P10:C1} decreases, while the other terms remain unchanged, thus \eqref{P10:C1} holds.

Next, we show that $\tilde{b}_{j,t} \le {b}_{j,t}^\star$ and therefore, \eqref{P10:C2} holds and the resulting objective value is lower. In the second case, $\tilde{b}_{j,t} = 0$ and hence $\tilde{b}_{j,t} \le {b}_{j,t}^\star$. Thus, we only consider the first case: 
$\tilde{b}_{j,t} =  \frac{2K_j v_{j,t}}{|\mathbf{f}_t^H \mathbf{h}_{j,t}|}- b_{j,t}^\star \overset{\mathrm{(a)}}{\le} \frac{K_j v_{j,t}}{|\mathbf{f}_t^H \mathbf{h}_{j,t}|}  \overset{\mathrm{(b)}}{\le} {b}_{j,t}^\star,$
where (a) and (b) follow from the assumption that $\frac{K_j v_{j,t}}{|\mathbf{f}_t^H \mathbf{h}_{j,t}|} \le b_{j,t}^\star.$ 

Using the relation in \eqref{Proposition4-1}, we can restrict the feasible set of \eqref{P10} by adding  \eqref{Proposition4-1} as a constraint to the problem and remove the absolute values on the LHS of \eqref{P10:C1}, and we arrive at the equivalent problem \eqref{P11}.
\endIEEEproof

\section{Proof of Proposition 6}\label{Appendix:G}
\IEEEproof
Define $ \hat{\mathbf{u}}_t \triangleq \frac{1}{K} \sum_{m=1}^M \Big( \frac{\mathcal{R}[\mathbf{f}^{H}_t  \hat{\mathbf{h}}_{m, t} a_{m,t}]}{v_{m,t}}- K_m\Big)\mathbf{g}_{m,t} $, $ \tilde{\mathbf{u}}_t \triangleq \frac{1}{K} \sum_{m=1}^M\frac{\mathcal{R}[\mathbf{f}^{H}_t \tilde{\mathbf{h}}_{m,t} a_{m,t}]}{v_{m,t}}\mathbf{g}_{m,t}$, and $\tilde{u}_t = \frac{\sqrt{D}}{K} \sum_{m=1}^M \Big|K_m v_{m,t}- \mathcal{R}[\mathbf{f}_t^H \hat{\mathbf{h}}_{m, t} a_{m,t}]\Big| $ . Based on the definition of $\mathbf{e}_t$ in \eqref{ErrorDef-ICSI}, we have
\begin{align}
\hspace{-1 em}\mathbb{E}[&\|\mathbf{e}_t\|^2| \mathbf{w}_t]  \overset{\mathrm{(a)}}{=} \| \hat{\mathbf{u}}_t  \|^2 + \mathbb{E}\Big[ \Big\| \frac{\tilde{\mathbf{n}}_t}{K}  \Big\|^2 \Big| \mathbf{w}_t \Big] + \mathbb{E}\Big[\| \tilde{\mathbf{u}}_t\|^2 \big| \mathbf{w}_t\Big]
\nonumber \\ & \overset{\mathrm{(b)}}{\le} \tilde{u}_t^2+ \frac{ D \sigma_n^2 \| \mathbf{f}_t\|^2}{ 2K^2}  + \mathbb{E}\Big[ \|  \tilde{\mathbf{u}}_t \|^2\big| \mathbf{w}_t\Big] \nonumber \\ &  \overset{\mathrm{(c)}}{=} \tilde{u}_t^2+ \frac{ D \sigma_n^2 \| \mathbf{f}_t\|^2}{ 2K^2} + \frac{D}{K^2} \sum_{m=1}^M\ \mathbb{E}\big|\mathcal{R}[\mathbf{f}^{H}_t \tilde{\mathbf{h}}_{m,t} a_{m,t}]\big|^2  \nonumber \\ & \overset{\mathrm{(d)}}{=} 
\tilde{u}_t^2  +  \frac{ D \sigma_n^2 \| \mathbf{f}_t\|^2}{ 2K^2} + \frac{ D \epsilon  \| \mathbf{f}_t\|^2}{ 2K^2} \sum_{m=1}^M \sigma_{h_m}^2 \left|a_{m,t}\right|^2,
\end{align}
where (a) is established by expanding the expression for $\|\mathbf{e}_t\|^2$ and leveraging the independence and zero-mean properties of $\tilde{\mathbf{h}}_{m,t}$ and $\mathbf{n}_{d,t}$. Inequality (b) follows by applying the Triangle Inequality to the first term and replacing $\mathbb{E}[\|\tilde{\mathbf{n}}_t\|^2]$ in the second term using the result in Appendix~\ref{Appendix:C}. 
Equality (c) results from expanding the term within the expectation and accounting for the independence of channel estimation errors 
$\tilde{\mathbf{h}}_{m,t}$ across devices, as well as their independence from $\mathbf{w}_t$. Equality (d) follows from the fact that $\mathbb{E}\big| \mathcal{R}[\mathbf{f}^{H}_t \tilde{\mathbf{h}}_{m,t} a_{m,t}]\big|^2 = \frac{\epsilon \sigma_{h_m}^2\|\mathbf{f}_t\|^2 \left|a_{m,t}\right|^2}{2}$ since $\tilde{\mathbf{h}}_{m,t} \sim \mathcal{CN}(\mathbf{0}, \epsilon \sigma_{h_m}^2 \mathbf{I})$.

\endIEEEproof

\section{Proof of Proposition 
7}\label{Appendix:H}
\IEEEproof
The proof is similar to the proof of Proposition \ref{Proposition4}. We can establish that \eqref{Proposition4-1} holds for the optimal solution of \eqref{ImperfectCSIProblem2}. Consequently, we can incorporate \eqref{Proposition4-1} as a constraint in the problem to restrict the feasible set and remove the absolute values on the LHSs of \eqref{ImperfectCSIProblem2:c1} and \eqref{ImperfectCSIProblem2:c2}. This leads to the equivalent problem \eqref{ImperfectCSIProblem3}.
\endIEEEproof

\bibliographystyle{IEEEbib}
\bibliography{Refs}

\end{document}